\documentclass[a4paper,11pt]{article}
\pdfoutput=1 % submitting a pdflatex build (images in pdf/png/jpg)

\usepackage{jcappub} % for details see the JCAP-author-manual
\usepackage[T1]{fontenc}

\usepackage{array}      % extended column specifiers
\usepackage{multirow}   % grouped row labels in the results table
\usepackage{xcolor}     % coloured TODO flags (remove before final submission)

\def\msun{h^{-1}{M_\odot}}
\def\gpc{h^{-1}{\rm Gpc}}
\def\mpc{h^{-1}{\rm Mpc}}

\def\hmpci{h{\rm Mpc}^{-1}}

\title{The Environmental Dependence of Halo Intrinsic Alignments: Stronger Signals in Underdense Regions}

\author[a,b]{Masaya Ichikawa,}
\author[b,c]{Jingjing Shi,}
\author[d]{Toshiki Kurita,}
\author[b,c]{Masahiro Takada,}
\author[c,e,f,g]{Takahiro Nishimichi,}
\author[b,c]{and Linda Blot}

\affiliation[a]{Advanced Energy, Graduate School of Frontier Sciences, The University of Tokyo, Kashiwa, Chiba 277-8561, Japan}
\affiliation[b]{Center for Data-Driven Discovery (CD3), Kavli IPMU (WPI), UTIAS, The University of Tokyo, Kashiwa, Chiba 277-8583, Japan}
\affiliation[c]{Kavli Institute for the Physics and Mathematics of the Universe (WPI), The University of Tokyo Institutes for Advanced Study (UTIAS), The University of Tokyo, 5-1-5 Kashiwanoha, Kashiwa-shi, Chiba 277-8583, Japan}
\affiliation[d]{Max-Planck-Institut f\"ur Astrophysik, Karl-Schwarzschild-Str. 1, 85748 Garching, Germany}
\affiliation[e]{Department of Astrophysics and Atmospheric Sciences, Faculty of Science, Kyoto Sangyo University, Motoyama, Kamigamo, Kita-ku, Kyoto 603-8555, Japan}
\affiliation[f]{Center for Gravitational Physics and Quantum Information, Yukawa Institute for Theoretical Physics, Kyoto University, Kyoto 606-8502, Japan}
\affiliation[g]{RIKEN Center for Advanced Intelligence Project, 1-4-1 Nihonbashi, Chuo-ku, Tokyo 103-0027, Japan}

\emailAdd{ichikawa.masaya24@ae.k.u-tokyo.ac.jp}
\emailAdd{jshi@g.ecc.u-tokyo.ac.jp}

\abstract{The intrinsic alignment (IA) of galaxies and the dark matter haloes that host them is one of the leading astrophysical systematics for weak-lensing cosmology, yet how the IA signal depends on the large-scale environment in which haloes reside is not yet fully characterised. We use the high-resolution $N$-body simulations of the \textsc{Dark Quest} suite to measure the environmental dependence of the IA of dark matter haloes over the redshift range $z=0.1$--$1.5$. We quantify each halo's environment through the overdensity $\delta_8$, defined from the number of neighbouring haloes within $8\,\mpc$, and we isolate the environmental effect from its degeneracy with the halo-mass dependence by comparing the most overdense and most underdense haloes constructed to share the same halo-mass distribution. We find that haloes in underdense environments exhibit systematically larger IA amplitudes $A_{\rm IA}$ than haloes of the same mass in overdense environments, by a factor of $\sim1.5$--$1.8$, and that this trend persists across the mass and redshift ranges probed, strengthening towards low redshift. Using an orientation-only (unit-ellipticity) estimator, we further show that this environmental contrast is driven by a combination of two effects: haloes in underdense regions are both intrinsically less spherical and more strongly aligned with the large-scale tidal field than their overdense counterparts of the same mass. These results indicate that the large-scale environment is a non-negligible variable in modelling halo and galaxy alignments, and may be a particularly important factor for beyond-two-point weak-lensing analyses.}


\begin{document}
\maketitle
\flushbottom

%%%%%%%%%%%%%%%%%%%%%%%%%%%%%%%%%%%%%%%%%%%%%%%%%%

%%%%%%%%%%%%%%%%% BODY OF PAPER %%%%%%%%%%%%%%%%%%

%%%%%%%%%%%%%%%%%%%%%%%%%%
%Introduction
%%%%%%%%%%%%%%%%%%%%%%%%%%
\section{Introduction}

Weak gravitational lensing \citep{Bartelmann_Schneider_2001,Hoekstra_Jain_2008,Kilbinger_2015} is one of the most powerful probes in modern cosmology because it is directly sensitive to the total matter distribution, including both baryons and dark matter. By measuring the coherent distortions imprinted on the observed shapes of distant galaxies by the intervening tidal field, weak-lensing surveys constrain the growth of cosmic structure and the expansion history of the Universe.

Realising the full statistical power of weak lensing, however, requires that several astrophysical and observational systematics be controlled to high precision, especially in the non-linear regime where linear theory breaks down and baryonic physics becomes important. Among these, the intrinsic alignment (IA) of galaxy shapes is one of the most important effects\citep{PhysRevD.70.063526,2007NJPh....9..444B, 2015PhR...558....1T,2015SSRv..193....1J,Lamman_2024,Chisari_2025}. Because galaxies and their host haloes form and evolve within a common large-scale tidal field, their shapes acquire physically correlated orientations that are not of lensing origin; if neglected, these correlations contaminate the measured cosmic-shear signal and bias the inferred cosmological parameters \citep{Krause_2016}.

A range of analytic models has been developed over the past two decades. The simplest is the linear alignment (LA) model \citep{Catelan_2001,PhysRevD.70.063526}, in which galaxy and halo shapes respond linearly to the large-scale gravitational tidal field; this model is expected to be valid only on large scales ($k \lesssim 0.1\,\hmpci$). To extend the description into the mildly non-linear regime, the non-linear alignment (NLA) model replaces the linear matter power spectrum with its non-linear counterpart \citep{2007NJPh....9..444B,PhysRevD.70.063526}, and more recently the tidal alignment and tidal torquing (TATT) model \citep{PhysRevD.100.103506} has added quadratic (tidal-torquing) contributions and density weighting, providing a more flexible description on smaller scales 
(see also \cite{Vlah+2020:IA_EFT,Vlah+2021:IA_EFT,Bakx+2023:EFTofIAvsSims,Chen&Kokron2024:LEFTofIA} for recent effective-field-theory-based models).
All of these models contain at least one free amplitude parameter, the IA amplitude $A_{\rm IA}$, whose value depends on the properties of the galaxy or halo sample under consideration, and a more accurate determination of $A_{\rm IA}$ for a given sample translates directly into tighter and less biased cosmological constraints. Observationally, $A_{\rm IA}$ is found to depend strongly on galaxy colour, luminosity, and redshift, with red, luminous, early-type galaxies exhibiting the strongest alignments \citep{2015MNRAS.450.2195S,2019A&A...624A..30J,Siegel_2025}.

These models nevertheless face important challenges. A central one is their validation for beyond-two-point statistics, which encode the non-Gaussian information of the matter field and are increasingly exploited to sharpen cosmological constraints. The widely used NLA model parametrises the IA amplitude of a given galaxy sample as a function of luminosity (or mass) and redshift only, and has mainly been validated for two-point statistics up to mildly non-linear scales. Accurately modelling IA at the level of higher-order statistics -- such as the alignment bispectrum and the full three-dimensional shape field \citep{2021JCAP...05..061V,2022PhRvD.105l3501K,Akitsu_2023} -- requires a more complete understanding of how halo and galaxy shapes respond, non-linearly, to their surrounding density and tidal fields.
We would also like to point out that IA effects are not merely a systematic contamination to weak lensing cosmology, but can also be used as a cosmological probe~\citep{2021PhRvD.103h3508A,2023PhRvD.108h3533K,2023ApJ...945L..30O,Kurita_2026}.

There is growing evidence that the large-scale environment affects the IA signal in addition to halo mass. Shi et al.~\cite{Shi_2015} found that the alignment of halo spins with the large-scale tidal field is stronger in weaker (more underdense) tidal environments. Using cosmological simulations, Xia et al.~\cite{2017ApJ...848...22X} showed that halo alignment depends not only on mass but also differs between haloes of the same mass in clusters versus filaments. Akitsu et al.~\cite{Akitsu_2021} demonstrated that the response of halo shapes to a large-scale tide depends on halo assembly history and axis ratio -- a ``shape assembly bias'' -- so that haloes of the same mass but with lower concentration and lower axis ratio (more aspherical) align more strongly. On the theoretical and observational side, Reischke \& Sch\"afer~\cite{2019JCAP...04..031R} predicted an environmental dependence of the IA ellipticity correlation functions, while d'Assignies D.\ et al.~\cite{2022MNRAS.509.1985D} measured coherent alignments of galaxies around cosmic voids. Together these works suggest that the IA amplitude inferred for a given sample may depend systematically on the density of the environments it occupies, even at fixed halo mass.

Motivated by this, we use the high-resolution $N$-body simulations of the \textsc{Dark Quest} suite~\citep{Nishimichi_2019} to measure the three-dimensional IA power spectrum and isolate the environmental dependence from the halo-mass dependence, by comparing the most overdense and most underdense haloes constructed to share the same mass distribution. We aim to answer whether, at fixed mass, haloes in underdense environments are more strongly aligned than those in overdense environments, and if so whether this arises from a genuine difference in alignment or simply from a difference in halo shape. The remainder of this paper is organised as follows. In Section~\ref{sec:data} we describe the simulations, our definition of the halo environment, the shape measurement, and the IA power-spectrum estimator. In Section~\ref{sec:results} we present the measured $P_{\delta E}$ and $P_{EE}$ spectra and the fitted IA amplitudes as functions of environment, mass, and redshift, and disentangle the shape and alignment contributions. We summarise our findings and discuss their implications in Section~\ref{sec:summary}.

%%%%%%%%%%%%%%%%%%%%%%%%%%
%Data and Measurement
%%%%%%%%%%%%%%%%%%%%%%%%%%
\section{Data and Measurement}
\label{sec:data}

\subsection{Simulation Data}

In this study we use the dark matter halo catalogues generated by the \textsc{Dark Quest} $N$-body simulation suite \citep{Nishimichi_2019}. The simulations follow $2048^3$ particles in comoving cubes with a side length of $1\,\gpc$ (high-resolution runs, HR). The suite consists of fixed-cosmology ``fiducial'' runs and varied cosmology runs within the $w$CDM cosmologies, with the former used to calibrate cosmic variance, and the latter for emulator construction.
% The simulation adopts the fiducial 
Here, we focus on the fiducial runs for the
$\Lambda$CDM cosmology: $(\omega_b, \omega_c, \Omega_{\rm de}, \ln(10^{10}A_s), n_s, w) = (0.02225, 0.1198, 0.6844, 3.094, 0.9645, -1)$, which corresponds to the best-fitting cosmological parameters from the \textit{Planck} 2015 \citep{refId0}. For the initial conditions, the linear matter power spectrum is computed from these parameters with \textsc{camb} \citep{Lewis_2000}; a Gaussian random field is generated from this spectrum, and the particle displacements and velocities are assigned using second-order Lagrangian perturbation theory \citep[2LPT;][]{10.1046/j.1365-8711.1998.01845.x,10.1111/j.1365-2966.2006.11040.x}. The subsequent gravitational evolution is computed with the parallel Tree--Particle-Mesh code \textsc{gadget-2} \citep{10.1111/j.1365-2966.2005.09655.x}. The mass of the N-body particles is $1.02\times 10^{10}\msun$.

Dark matter haloes are identified in post-processing with the phase-space halo finder \textsc{rockstar} \citep{Behroozi_2013}. \textsc{rockstar} provides the centre position of each halo together with several mass definitions; throughout this work we adopt the virial mass $M_h$ as the halo mass. We restrict our analysis to haloes with $M_h > 10^{12}\,\msun$ and consider six output redshifts in the range $z=0.1$--$1.5$.

\subsection{Environment Measurement}

\begin{figure}
\begin{center}
	\includegraphics[width=0.6\columnwidth]{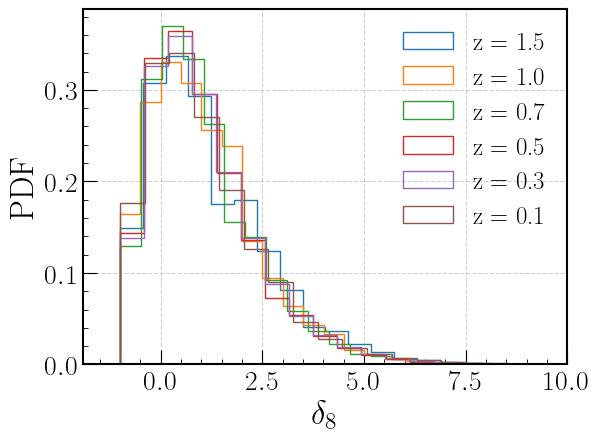}
    \caption{Probability distribution function of the large-scale overdensity $\delta_8$ for haloes with $M_{h}>10^{12}\,\msun$ at six redshifts between $z=0.1$ and $z=1.5$. The distribution of $\delta_8$ is nearly independent of redshift.}
    \label{fig:pdf_delta8}
\end{center}
\end{figure}

To study the relationship between the intrinsic alignment of dark matter haloes and their large-scale environment, we characterise the environment of each halo by the number of neighbouring haloes, $N_8$, within a sphere of radius $8\,\mpc$, and define the corresponding overdensity
\begin{equation}
    \delta_8 \equiv \frac{N_8}{\langle N_8 \rangle} - 1,
	\label{eq:delta8}
\end{equation}
where $\langle N_8 \rangle$ is the mean number of haloes expected in the same volume. The smoothing scale of $8\,\mpc$ is large enough that $\delta_8$ traces the quasi-linear large-scale environment while remaining a statistically robust tracer of the local density \citep{Abbas_2007}. Figure~\ref{fig:pdf_delta8} shows the probability distribution function of $\delta_8$ at each redshift; the distribution is nearly independent of redshift over the range considered. The distribution of $\delta_8$ can be reasonably described even by Poisson cluster models (see Appendix of \cite{Abbas_2007}).

\begin{figure}
\begin{center}
	\includegraphics[width=0.8\linewidth]{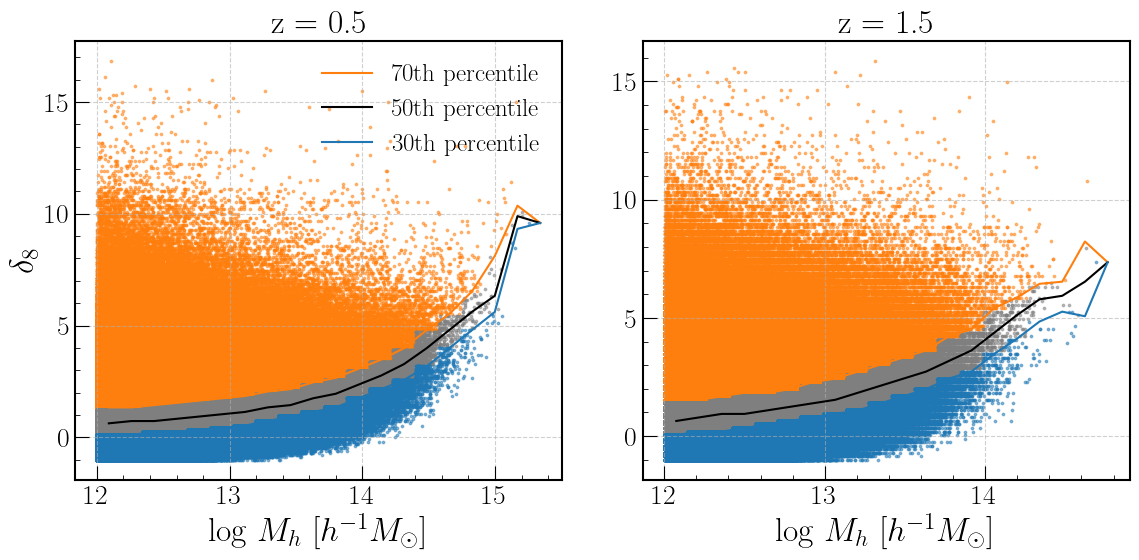}
    \caption{Large-scale overdensity $\delta_8$ as a function of halo mass $M_{h}$ at $z=0.5$ (left) and $z=1.5$ (right). Orange and blue points denote haloes classified as overdense and underdense, respectively (the upper and lower $30$ per cent of $\delta_8$ within each mass bin); grey points show the remaining haloes. The three solid lines show the $30$th, $50$th, and $70$th percentiles of $\delta_8$ in each mass bin. The median overdensity increases with halo mass, motivating the mass-binned classification of environment.}
    \label{fig:delta8_mhalo}
\end{center}
\end{figure}

Because halo mass is itself one of the strongest drivers of the intrinsic alignment, a naive classification based on $\delta_8$, would conflate the environmental dependence with the mass dependence: massive halos tend to live in overdense regions (Figure~\ref{fig:delta8_mhalo}). To break this degeneracy, we instead divide the haloes into $20$ even-logarithmic mass bins and, within each bin, classify the upper $30$ per cent in $\delta_8$ as the overdense subsample and the lower $30$ per cent as the underdense subsample. This procedure matches the mass distributions of the two subsamples by construction, so that any residual difference in their alignment signals can be attributed to environment rather than to mass. Figure~\ref{fig:delta8_mhalo} shows the distribution of haloes in the $\delta_8$--$M_h$ plane: orange and blue points mark the overdense and underdense subsamples, respectively, and grey points the remaining haloes. The median $\delta_8$ (the $50$th-percentile line) rises with halo mass, confirming that more massive haloes preferentially inhabit denser environments.

\subsection{Shape Measurement}

To study the intrinsic alignment of dark matter haloes we must first quantify the shape of each individual halo. Since a halo is not a rigid body but a self-gravitating collection of particles, there is no unique definition of its shape. In Appendix~C of Kurita et al.~\cite{Kurita_2020}, the IA power spectra obtained from different definitions of the halo inertia tensor were compared. While the spectra from different inertia-tensor definitions differ by overall constant factors, on large scales ($k \lesssim 0.1\,\hmpci$) they share the same $k$-dependence and yield a nearly constant signal-to-noise ratio. Consequently, on the large scales of interest here, essentially the same information is recovered regardless of the inertia-tensor definition. We therefore adopt the default definition of Kurita et al.~\cite{Kurita_2020}, namely the reduced inertia tensor computed with the iterative method \citep[see also][]{Osato_2018,Shi_2021_JCAP}, which reduces the sensitivity to an arbitrary choice of boundary radius,
\begin{equation}
    I_{ij} \equiv \frac{1}{\sum_p m_p}\sum_p m_p\frac{\Delta x_p^i \, \Delta x_p^j}{r_p^2},
	\label{eq:inertia}
\end{equation}
where $\Delta \boldsymbol{x}_p \equiv \boldsymbol{x}_p - \boldsymbol{x}_h$ is the position of particle $p$ relative to the halo centre $\boldsymbol{x}_h$, and the indices $i,j$ run over the two dimensions of the plane perpendicular to the line of sight, which is the direction of $x_3$ in this work. The weight $1/r_p^2$ suppresses the contribution of particles in the outer regions of the halo. In the iterative scheme, $r_p$ is initially set to the virial radius and is thereafter updated as
\begin{equation}
    r_p \equiv \sqrt{(\boldsymbol{x}_p \cdot \boldsymbol{e}_a)^2 + \frac{(\boldsymbol{x}_p \cdot \boldsymbol{e}_b)^2}{s^2} + \frac{(\boldsymbol{x}_p \cdot \boldsymbol{e}_c)^2}{q^2}},
	\label{eq:rp}
\end{equation}
where $\boldsymbol{e}_a, \boldsymbol{e}_b, \boldsymbol{e}_c$ are the eigenvectors of the inertia tensor obtained in the previous iteration (ordered such that $a>b>c$), and $q \equiv c/a$ and $s \equiv b/a$ are the axis ratios. The iteration is terminated once $q$ and $s$ have converged to within $1$ per cent.

We then define the two projected ellipticity components of the halo from the inertia tensor as
\begin{gather}
    e_+ \equiv \frac{I_{11} - I_{22}}{I_{11} + I_{22}}, \\
    e_\times \equiv \frac{2 I_{12}}{I_{11} + I_{22}}.
	\label{eq:ellipticity}
\end{gather}
This definition is analogous to the galaxy ellipticity measured from the quadrupole moments of the surface-brightness distribution in weak-lensing observations. To compare directly with the IA model described in the next subsection, we convert the halo ellipticity to a shear using the standard weak-lensing relation \citep{2002AJ....123..583B}
\begin{equation}
    \gamma_{+,\times} = \frac{e_{+,\times}}{2\mathcal{R}},
	\label{eq:shear}
\end{equation}
where $\mathcal{R} \equiv 1 - \langle e_i^2 \rangle$ is the shear responsivity and $\langle e_i^2 \rangle$ is the ellipticity variance per component for a given galaxy sample. We evaluate $\mathcal{R}$ separately for each subsample.
Note that the shear field estimated from halo shapes is a density-weighted field because the shape field can be sampled only at the positions of halos~\citep[see][for a similar discussion]{Kurita_2020}.

We emphasise that the responsivity normalisation rescales the ellipticity into a shear-equivalent amplitude but does not remove the dependence of the measured signal on the intrinsic ellipticity magnitude: the IA power spectra are sensitive to the product of the typical halo elongation and the degree to which halo orientations are aligned with the tidal field. An environmental dependence of the IA spectra could therefore arise either from a genuine difference in the alignment of halo orientations, or simply from a difference in how elongated the haloes are. We disentangle these two contributions in Section~\ref{sec:shape_vs_align} by comparing the ellipticity of the over- and underdense subsamples directly.

\subsection{IA Power Spectrum: measurement and modelling}
\label{sec:ia_ps}

\subsubsection{Measurement}
\label{sec:ia_meas}

The halo shear field $\gamma_{(+,\times)}$ defined in equation~(\ref{eq:shear}) is a spin-$2$ field, which we decompose into $E$- (curl-free) and $B$- (divergence-free) modes in Fourier space, in analogy with cosmic microwave background polarisation and weak lensing \citep{Kurita_2020,Shi_2021_JCAP},
\begin{gather}
\label{eq:eb}
    \gamma_E(\boldsymbol{k}) = \gamma_{+}(\boldsymbol{k}) \cos{2\phi_k} + \gamma_{\times}(\boldsymbol{k}) \sin{2\phi_k}, \\
    \gamma_B(\boldsymbol{k}) = -\gamma_{+}(\boldsymbol{k}) \sin{2\phi_k} + \gamma_{\times}(\boldsymbol{k}) \cos{2\phi_k},
\end{gather}
where the wavevector is $\boldsymbol{k} = k(\sqrt{1-\mu^2}\cos\phi_k,\, \sqrt{1-\mu^2}\sin\phi_k,\, \mu)$, with $\mu$ the cosine of the angle between $\boldsymbol{k}$ and the line of sight and $\phi_k$ the azimuthal angle. We measure the cross-power spectrum between the $E$-mode shear and the matter density field, $P_{\delta E}$, and the $E$-mode auto-power spectrum, $P_{EE}$, defined by
\begin{gather}
\label{eq:ps_def}
    \langle \gamma_E(\boldsymbol{k})\, \delta(\boldsymbol{k}^{\prime}) \rangle = (2\pi)^3\, \delta_D(\boldsymbol{k} + \boldsymbol{k}^{\prime})\, P_{\delta E}(\boldsymbol{k}), \\
    \langle \gamma_E(\boldsymbol{k})\, \gamma_E(\boldsymbol{k}^{\prime}) \rangle = (2\pi)^3\, \delta_D(\boldsymbol{k} + \boldsymbol{k}^{\prime})\, P_{EE}(\boldsymbol{k}).
\end{gather}
By parity the $B$-mode auto-spectrum $P_{BB}$ vanishes in the linear regime, receiving contributions only from the shape (shot) noise and weak non-linear effects; we exploit this to estimate the noise (Section~\ref{sec:results} and Appendix~\ref{app:extra}). Since the spectra depend on both $k$ and $\mu$, we characterise them through their multipole moments,
\begin{equation}
    P^{(\ell)}(k) = \frac{2\ell+1}{2} \int_{-1}^{1} d\mu\; \mathcal{L}_\ell(\mu)\, P(k, \mu),
	\label{eq:multipole}
\end{equation}
with $\mathcal{L}_\ell$ the Legendre polynomials; in this work we use the monopole ($\ell=0$).

In practice we assign the halo number density and the two shear components $\gamma_{+,\times}$ to a $1024^3$ Cartesian grid using cloud-in-cell interpolation, Fourier transform the fields, construct the $E$- and $B$-mode fields via equations~(\ref{eq:eb}), and average the products in equations~(\ref{eq:ps_def}) over spherical $k$-shells.

\subsubsection{Modelling: the (non-)linear alignment model}
\label{sec:ia_model}

To interpret the measured spectra we adopt the linear alignment (LA) model \citep{Catelan_2001,PhysRevD.70.063526}, in which the shape of a halo (or galaxy) responds linearly to the large-scale tidal gravitational field,
\begin{equation}
    \gamma^I_{(+, \times)} = -\frac{C_1}{4 \pi G} \left(\nabla_x^2 - \nabla_y^2,\; 2 \nabla_x \nabla_y\right) S[\Psi_p],
	\label{eq:la_real}
\end{equation}
where $C_1$ is a normalisation constant, $S$ is a smoothing filter that retains only large-scale modes, and $\Psi_p$ is the primordial gravitational potential, related to the density field through the Poisson equation,
\begin{equation}
    \Psi_p(\boldsymbol{k}, z_{\rm IA}) = -4 \pi G\, \bar{\rho}_{m0}\, (1 + z_{\rm IA})\, k^{-2}\, \delta(\boldsymbol{k}, z_{\rm IA}).
	\label{eq:poisson}
\end{equation}
In this picture, halo shapes are assumed to be established over the halo formation timescale in environments regulated by the long-wavelength primordial tidal field~\citep{PhysRevD.70.063526,Akitsu_2021}.
In Fourier space the intrinsic shear is then
\begin{equation}
    \gamma^I_{(+, \times)}(\boldsymbol{k}, z) = -A_{\rm IA}\, C_1\, \frac{\bar{\rho}_{m0}}{D(z)}\, f_{(+, \times)}\, \delta(\boldsymbol{k}, z),
	\label{eq:la_fourier}
\end{equation}
with $f_+ \equiv (1-\mu^2)\cos 2\phi_k$, $f_\times \equiv (1-\mu^2)\sin 2\phi_k$, $D(z)$ the linear growth factor normalised to $D(0)=1$, and $A_{\rm IA}$ the dimensionless IA amplitude into which the dependence of the alignment strength on sample properties is absorbed. The model then predicts the IA power spectra in terms of the matter power spectrum $P_{\delta\delta}(k,z)$ \citep{Kurita_2020},
\begin{gather}
\label{eq:ps_la}
    P_{\delta E}(k, \mu) = -A_{\rm IA}\, C_1 \rho_{\rm cr0}\, \frac{\Omega_m}{D(z)}\, (1 - \mu^2)\, P_{\delta\delta}(k, z), \\
    P_{EE}(k, \mu) = \left[A_{\rm IA}\, C_1 \rho_{\rm cr0}\, \frac{\Omega_m}{D(z)}\right]^2 (1 - \mu^2)^2\, P_{\delta\delta}(k, z),
\end{gather}
where $\rho_{\rm cr0}$ is the critical density today and $\bar{\rho}_{m0} = \Omega_m \rho_{\rm cr0}$; following common convention we adopt $C_1\rho_{\rm cr0} = 0.0134$ \citep{2007NJPh....9..444B}. The cross-spectrum $P_{\delta E}$ has angular modulation up to $\mu^2$ (non-zero monopole and quadrupole), whereas the auto-spectrum $P_{EE}$ varies up to $\mu^4$. The non-linear alignment (NLA) model extends this prescription to mildly non-linear scales by replacing the linear $P_{\delta\delta}$ with its non-linear counterpart \citep{2007NJPh....9..444B}. We fit $A_{\rm IA}$ from the measured monopole $P^{(0)}_{\delta E}$ relative to the matter power spectrum over $0.02 < k < 0.07\,\hmpci$, as described in Section~\ref{sec:results}.

%%%%%%%%%%%%%%%%%%%%%%%%%%
%Results
%%%%%%%%%%%%%%%%%%%%%%%%%%
\section{Results}
\label{sec:results}

In this section we present our measurements of the environmental dependence of halo intrinsic alignments. We begin in Section~\ref{sec:results_main} with two broad mass ranges, $10^{12} < M_h < 10^{13}\,\msun$ and $10^{13} < M_h < 10^{14}\,\msun$, at $z=0.5$ and $z=1.5$, comparing the overdense and underdense subsamples defined in Section~\ref{sec:data}: we measure the cross-power spectrum between the $E$-mode shape field and the matter density field, $P_{\delta E}$ (Figure~\ref{fig:P_de}), and the $E$-mode auto-power spectrum, $P_{EE}$ (Figure~\ref{fig:P_ee}), and quantify the alignment strength through the fitted amplitude $A_{\rm IA}$. We then extend the analysis to four narrower mass bins and six redshifts from $z=1.5$ to $z=0.1$ (Figure~\ref{fig:redshift_mass_dependance}), mapping out the joint dependence of $A_{\rm IA}$ on environment, halo mass, and redshift. Finally, in Section~\ref{sec:shape_vs_align} we disentangle the two distinct contributions to the measured signal -- the intrinsic halo shape and the degree of alignment -- to identify which drives the environmental trend.

\subsection{Environmental dependence of halo IA}
\label{sec:results_main}

\begin{figure}
\begin{center}
	\includegraphics[width=0.8\linewidth]{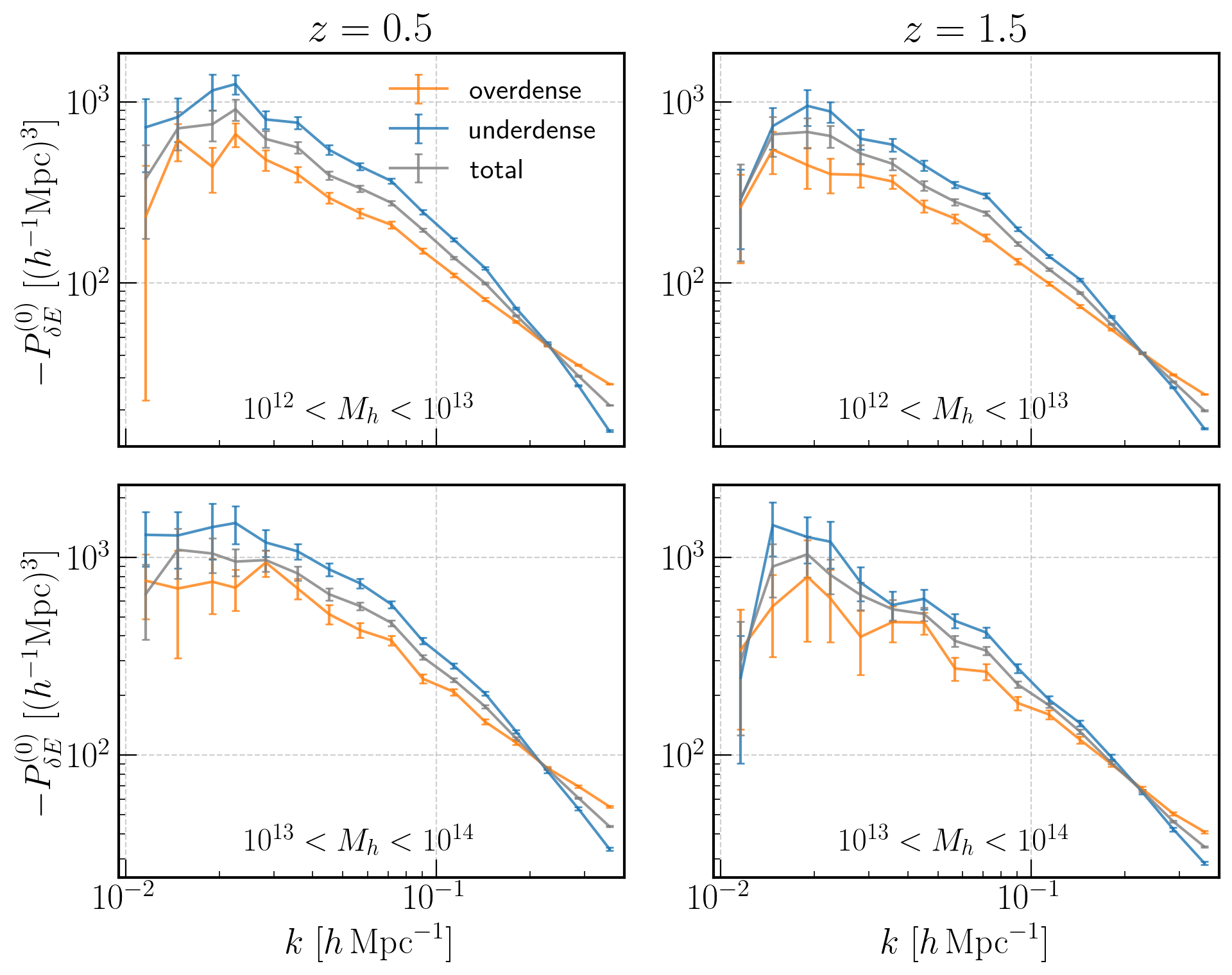}
    \caption{Cross-power spectrum between the $E$-mode halo shape field and the matter density field, $-P^{(0)}_{\delta E}(k)$, for the overdense (orange) and underdense (blue) subsamples. The columns correspond to $z=0.5$ (left) and $z=1.5$ (right), and the rows to the mass ranges $10^{12} < M_h < 10^{13}\,\msun$ (top) and $10^{13} < M_h < 10^{14}\,\msun$ (bottom). Error bars show the statistical uncertainty. On large scales the underdense subsample exhibits a stronger IA signal than the overdense subsample.}
    \label{fig:P_de}
\end{center}
\end{figure}

\begin{figure}
\begin{center}
	\includegraphics[width=0.85\linewidth]{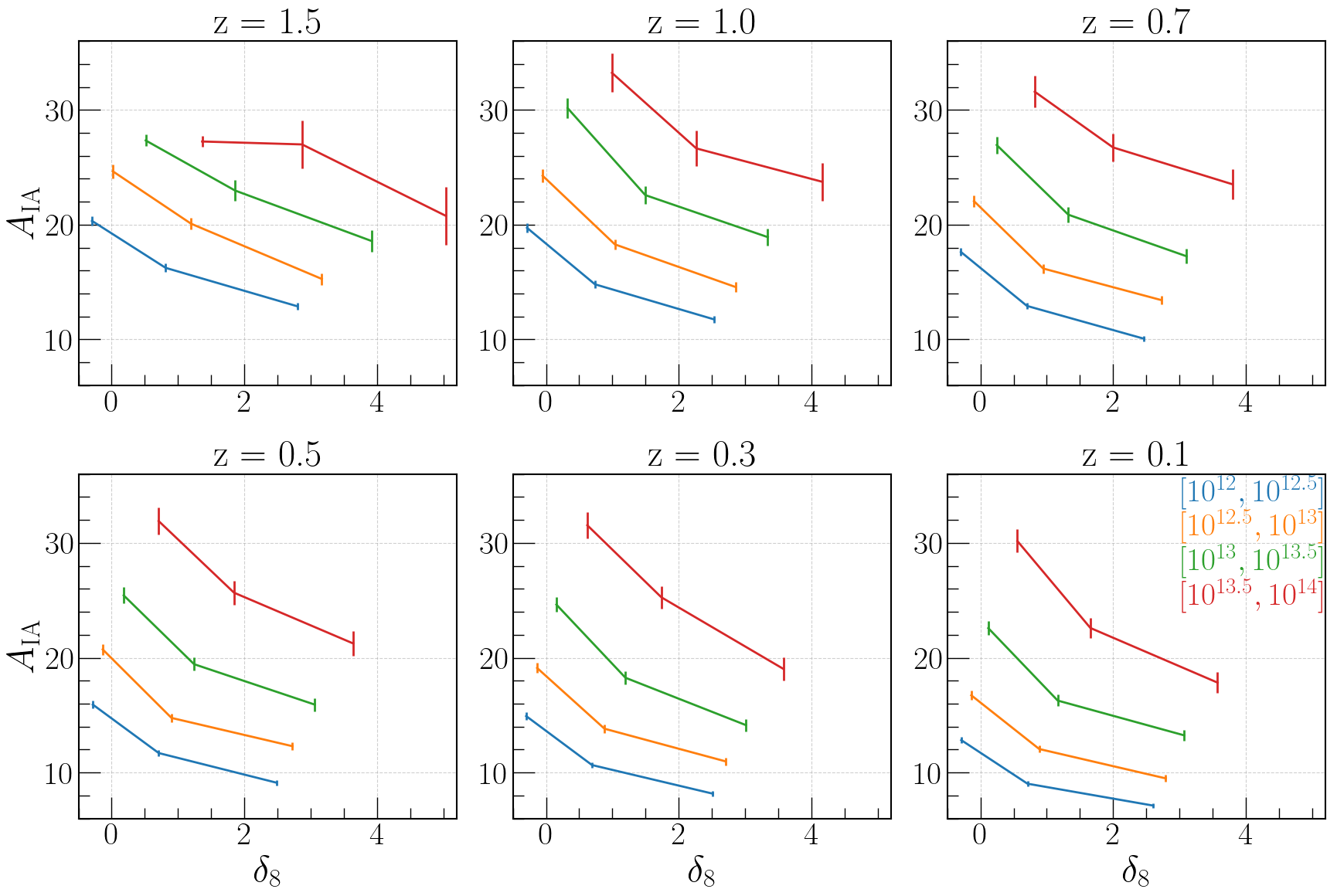}
    \caption{Best-fitting IA amplitude $A_{\rm IA}$ as a function of the mean large-scale overdensity $\delta_8$, in six redshift panels from $z=1.5$ to $z=0.1$. Within each panel, the coloured lines correspond to four halo mass bins, $10^{12}$--$10^{12.5}$, $10^{12.5}$--$10^{13}$, $10^{13}$--$10^{13.5}$, and $10^{13.5}$--$10^{14}\,\msun$. For a given redshift and halo mass bin, $\delta_8$ is calculated for the underdense, intermediate, and overdense subsamples separately. The IA amplitude decreases with increasing $\delta_8$ at fixed halo mass and increases with halo mass, across all redshifts. The environmental dependence is slightly stronger for lower halo mass bins. }
    \label{fig:redshift_mass_dependance}
\end{center}
\end{figure}

Figure~\ref{fig:P_de} shows the cross-power spectrum $-P^{(0)}_{\delta E}$ of the $E$-mode shape field and the matter density field. For both mass ranges, the underdense subsample shows a stronger IA signal than the overdense subsample at both redshifts at large scales, indicating that the shapes of haloes in underdense environments are more strongly correlated with the surrounding matter density field. On smaller scales the overdense subsample instead dominates, which mainly comes from our division of overdense versus underdense based on $\delta_8$. 
For the higher mass range, $10^{13} < M_h < 10^{14}\,\msun$, the measurement is noisier, especially at $z=1.5$,
owing to the small number of massive haloes. We fit the measured IA power spectrum in the range of $0.02 < k < 0.07\,\hmpci$, which is chosen to balance the regime of validity of the linear alignment model against the numerical and statistical limitations of the measurement. The lower bound, $k=0.02\,\hmpci$, lies a few times above the fundamental mode of the box, $k_{\rm f} = 2\pi/L \simeq 0.006\,\hmpci$, so that each $k$-bin contains enough independent modes to suppress the large sample variance that affects the very largest scales. The upper bound, $k=0.07\,\hmpci$, keeps the fit safely within the linear regime ($k \lesssim 0.1\,\hmpci$) where the linear alignment model applies; it is also far below the grid Nyquist wavenumber, $k_{\rm Ny} = \pi N_{\rm grid}/L \simeq 3.2\,\hmpci$, so that mass-assignment effects are negligible. We then quantify the alignment strength by fitting the IA amplitude $A_{\rm IA}$ to the measured spectra over this range using a $\chi^2$ minimisation, modelling the cross-spectrum as $P_{\delta E}^{(0)} = \tfrac{2}{3} b_K P_{\delta\delta}$ with $b_K = C_1\bar\rho_{\rm cr}\Omega_{\rm m}\,A_{\rm IA}/D(z)$, where $P_{\delta\delta}$ is the matter auto-power spectrum measured from the same simulation. The resulting best-fitting amplitudes for the overdense and underdense subsamples are summarised in Table~\ref{tab:aia}. At fixed mass and redshift the underdense subsample has a systematically larger $A_{\rm IA}$ than the overdense subsample, by a factor of $\sim1.5$--$1.8$. The low mass sample shows a slightly stronger environmental dependence, and such dependence seems slightly stronger at low $z$. We present a more quantative results in Figure~\ref{fig:redshift_mass_dependance}. We have also verified that this environmental trend is robust to the choice of halo shape estimator: repeating the measurement with an alternative, unreduced (iterative) inertia tensor yields the same result (Fig.~\ref{fig:P_de_SIAC}, Appendix~\ref{app:estimator}).

\begin{table}
\centering
\caption{Best-fitting linear-alignment amplitude $A_{\rm IA}$ for the overdense and underdense subsamples, from a least-$\chi^2$ fit of $P^{(0)}_{\delta E}$ against the matter auto-power spectrum over $0.02 < k < 0.07\,\hmpci$. Columns 5 and 6 give the underdense-to-overdense ratio of the IA amplitude for the standard and the orientation-only (normalised, $\hat{A}_{\rm IA}$) estimators. The following three columns give the per-component rms ellipticity $\langle e_i^2\rangle^{1/2}$ of the two subsamples and its underdense-to-overdense ratio; the standard ratio is well approximated by the product of the normalised (alignment) and the rms-ellipticity (shape) ratios. The last two columns give the linear halo bias $b_h = P_{\delta h}/P_{\delta\delta}$ of each subsample, fitted over the same $k$-range ($1\sigma$ fit errors).}
\label{tab:aia}
\resizebox{\textwidth}{!}{%
\begin{tabular}{l  c c c c c c c c c c}
\hline
$M_h$ $[\msun]$ & $z$ & $A_{\rm IA}^{\rm over}$ & $A_{\rm IA}^{\rm under}$ & $\dfrac{A_{\rm IA}^{\rm under}}{A_{\rm IA}^{\rm over}}$ & $\dfrac{\hat{A}_{\rm IA}^{\rm under}}{\hat{A}_{\rm IA}^{\rm over}}$ & $e_{\rm rms}^{\rm over}$ & $e_{\rm rms}^{\rm under}$ & $\dfrac{e_{\rm rms}^{\rm under}}{e_{\rm rms}^{\rm over}}$ & $b_h^{\rm over}$ & $b_h^{\rm under}$ \\
\hline
\multirow{2}{*}{$10^{12} - 10^{13}$} & $0.5$ & $9.88 \pm 0.43$  & $18.05 \pm 0.67$ & $1.83$ & $1.66$ & $0.319$ & $0.341$ & $1.07$ & $3.11 \pm 0.07$ & $-0.36 \pm 0.01$ \\
                                            & $1.5$ & $13.60 \pm 0.61$ & $21.90 \pm 0.82$ & $1.61$ & $1.54$ & $0.360$ & $0.376$ & $1.05$ & $4.73 \pm 0.11$ & $-0.30 \pm 0.02$ \\
\hline
\multirow{2}{*}{$10^{13} - 10^{14}$} & $0.5$ & $16.92 \pm 0.98$ & $28.32 \pm 1.25$ & $1.67$ & $1.53$ & $0.303$ & $0.326$ & $1.08$ & $3.77 \pm 0.09$ & $0.32 \pm 0.02$ \\
                                            & $1.5$ & $18.40 \pm 1.57$ & $28.12 \pm 1.73$ & $1.53$ & $1.42$ & $0.326$ & $0.347$ & $1.06$ & $6.25 \pm 0.16$ & $1.29 \pm 0.05$ \\
\hline
\end{tabular}%
}
\end{table}

\begin{figure}
\begin{center}
	\includegraphics[width=0.8\linewidth]{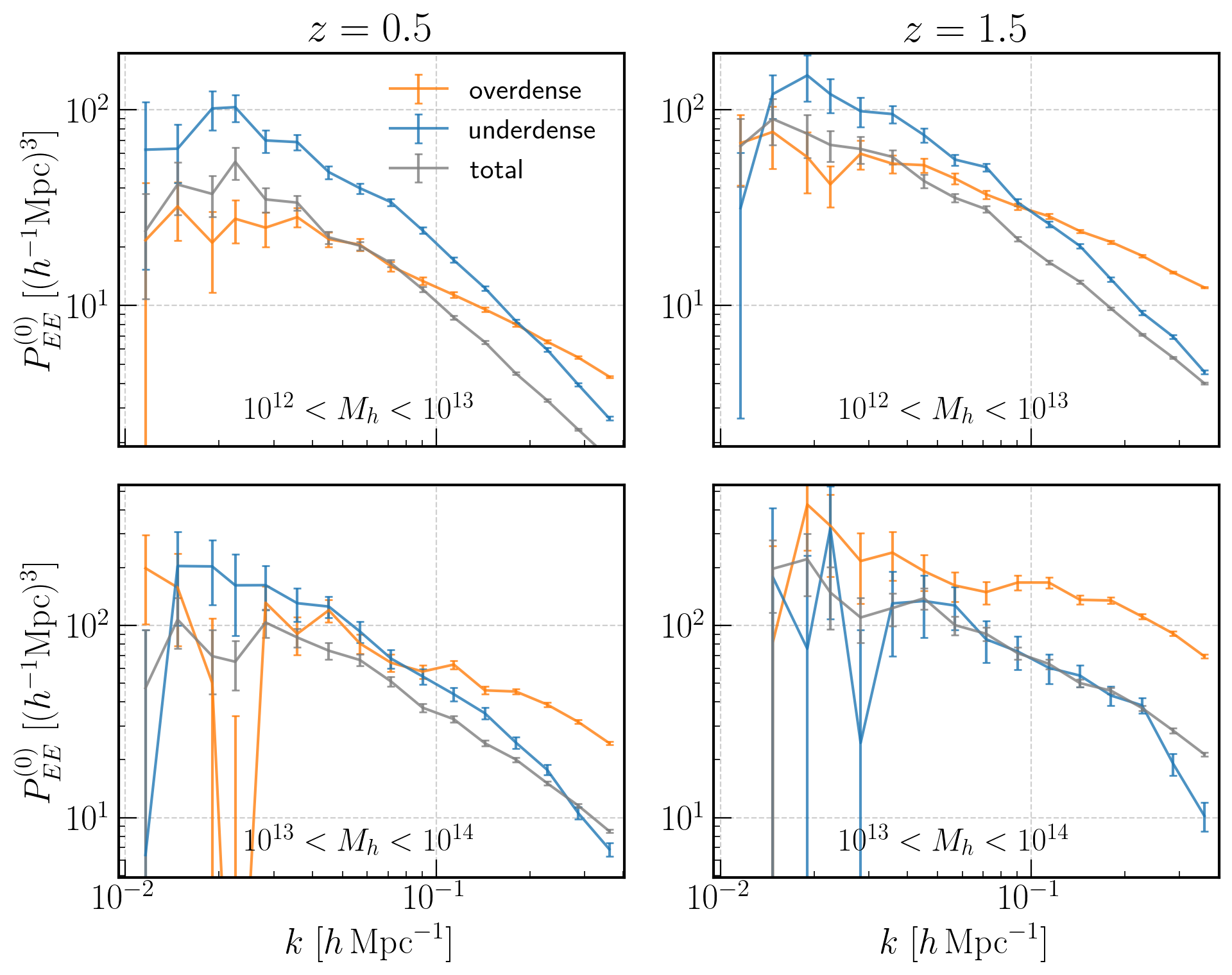}
    \caption{$E$-mode auto-power spectrum of the halo shape field, $P^{(0)}_{EE}(k)$, for the overdense (orange), underdense (blue), and total (grey) subsamples, with the analytic Poisson shape shot noise subtracted. The panels follow the same layout as Figure~\ref{fig:P_de}: $z=0.5$ (left) and $z=1.5$ (right), and mass ranges $10^{12} < M_h < 10^{13}\,\msun$ (top) and $10^{13} < M_h < 10^{14}\,\msun$ (bottom). On large scales the underdense subsample shows a stronger shape auto-correlation than the overdense subsample. The higher mass bin, and in particular its $z=1.5$ panel, is markedly noisier because of the small number of massive haloes (fewer still at high redshift), where the auto-spectrum becomes shot-noise dominated. The Poisson term does not fully capture the noise of the overdense sample, which carries an additional non-Poisson contribution; this is quantified in Appendix~\ref{app:extra}, where we also show the spectrum obtained by instead subtracting the measured $B$-mode (Figure~\ref{fig:P_ee_bmode}).}
    \label{fig:P_ee}
\end{center}
\end{figure}

Figure~\ref{fig:P_ee} shows the $E$-mode auto-power spectrum $P^{(0)}_{EE}$, which encodes the degree to which the orientations of neighbouring haloes are mutually aligned. Here we subtract the analytic Poisson shape shot noise, $P_{\rm shot} = \langle \sigma_\gamma^2 \rangle / \bar n_h$. Restricting attention to large scales, we find that the underdense subsample shows a stronger correlation than the overdense subsample, which is consistent with what was shown in Figure~\ref{fig:P_de}. However, we do notice the environmental effect is not as strong as expected from Figure~\ref{fig:P_de}, for the massive halos at $z=1.5$, the trend even seems reversed. 
In Appendix~\ref{app:extra}, we quantify the contribution from non-Poisson noise (Fig.~\ref{fig:pbb_resid}) using $B$-mode auto-power spectrum, which is particularly larger for the overdense subsample. 
In fact, once the non-Poisson noise is properly subtracted at large scale, the environmental effect strength looks consistent with Figure~\ref{fig:P_ee} (see Figure~\ref{fig:P_ee_bmode}).

The small-scale excess of the overdense $P_{\delta E}$ (Figure~\ref{fig:P_de}) and the enhanced overdense $P_{EE}$ can both be understood as a consequence of the \emph{density weighting} of the estimated shape field. Because halo shapes can be sampled only at halo positions, the measured shear is a density-weighted field, $\tilde{\boldsymbol\gamma}=(1+\delta_h)\,\boldsymbol\gamma$ \citep[see][]{Kurita_2020}, with $\delta_h=b_h\delta_m$ on large scales; the cross- and auto-spectra therefore pick up factors of $(1+\delta_h)$ and $(1+\delta_h)^2$, so in a region of matter overdensity $\delta_m$ they are boosted by roughly $(1+b_h\delta_m)$ and $(1+b_h\delta_m)^2$ respectively. The characteristic matter fluctuation, measured directly from the matter power spectrum as $\delta_m(k)=[k^3P_{\delta\delta}(k)/2\pi^2]^{1/2}$, is $\simeq0.05$ at $k=0.02\,\hmpci$ for the massive $z=1.5$ sample and grows to $\simeq0.2$ at $k=0.07\,\hmpci$ and $\simeq0.6$ at $k=0.3\,\hmpci$. Because the overdense subsample is strongly biased ($b_h^{\rm over}\simeq6.3$; Table~\ref{tab:aia}), $b_h\delta_m$ is not negligible even on large scales. For $P_{EE}$, which depends on $(1+\delta_h)^2$, the overdense amplitude is already boosted by $\simeq(1+b_h\delta_m)^2\simeq1.7$ at $k=0.02\,\hmpci$, so the density weighting is important for $P_{EE}$ even at the largest scales we fit. The apparent reversal of the massive overdense $P_{EE}$ (Figure~\ref{fig:P_ee}) is in fact driven by two distinct, clustering-driven contributions: (i) the residual \emph{non-Poisson noise} left after the analytic Poisson shot-noise subtraction, which can be removed by instead subtracting the measured $B$-mode (Appendix~\ref{app:extra}); and (ii) the $(1+\delta_h)^2$ boost of the signal itself, which the $B$-mode subtraction does not remove. Both originate from the strong clustering (high bias) of the overdense subsample, so that even after the non-Poisson noise is removed a residual signal-level boost remains, although it is sub-dominant to the intrinsic environmental contrast on the large scales of interest. For $P_{\delta E}$, which carries only a single power of $(1+\delta_h)$, the boost is much milder and remains sub-dominant across the fitting range ($k\lesssim0.07\,\hmpci$), so $P_{\delta E}$ preserves the underdense$\,>\,$overdense ordering there and the $A_{\rm IA}$ inferred from it is comparatively robust. Towards smaller scales, however, $\delta_m$ grows and $b_h\delta_m$ becomes of order unity or larger, so the boost becomes large for both $P_{\delta E}$ and $P_{EE}$ and strongly favours the more biased overdense sample; this drives the reversal of the environmental ordering at small scales ($k\sim0.3\,\hmpci$, where the overdense signal exceeds the underdense) seen in both $P_{\delta E}$ and $P_{EE}$, reflecting the halo clustering rather than a genuinely stronger intrinsic alignment.

We note, however, that part of the small-scale overdense excess could genuinely reflect a stronger one-halo intrinsic alignment in dense environments, such as the radial alignment of substructure \citep{VanAlfen_2024}. Disentangling this contribution from the density-weighting boost requires a more thorough analysis, which we leave to future work.

Figure~\ref{fig:redshift_mass_dependance} summarises the dependence of the fitted IA amplitude $A_{\rm IA}$ on environment, halo mass, and redshift. To resolve the mass dependence in more detail, here we subdivide the haloes into four narrower mass bins, $10^{12}$--$10^{12.5}$, $10^{12.5}$--$10^{13}$, $10^{13}$--$10^{13.5}$, and $10^{13.5}$--$10^{14}\,\msun$. At every redshift and in every mass bin, $A_{\rm IA}$ decreases monotonically with increasing $\delta_8$: haloes in underdense environments have systematically larger IA amplitudes than haloes of the same mass in overdense environments. $A_{\rm IA}$ increases with halo mass, recovering the well-established mass dependence of halo alignments. The environmental contrast also strengthens towards low redshift: for the lower-mass haloes the underdense-to-overdense amplitude ratio grows from $\sim1.6$ at $z=1.5$ to $\sim1.9$ at $z=0.1$, a difference detected at the $\sim8$--$10\sigma$ level, whereas for the higher-mass haloes the ratio stays close to $\sim1.6$ with little redshift evolution and is measured at lower significance ($\sim4$--$6\sigma$) because of the smaller halo numbers.

\subsection{Disentangling shape and alignment}
\label{sec:shape_vs_align}

As discussed in Section~\ref{sec:data}, the IA power spectra are sensitive to the product of the typical halo elongation and the alignment of halo orientations with the tidal field, and the responsivity normalisation does not remove the dependence on the intrinsic ellipticity magnitude. The stronger IA signal of the underdense haloes seen in Figures~\ref{fig:P_de}--\ref{fig:redshift_mass_dependance} could therefore reflect a difference in their intrinsic shapes, rather than, or in addition to, a difference in their alignment. To separate these two effects we compare the root-mean-square ellipticity, $\langle e_i^2 \rangle^{1/2}$, of the over- and underdense subsamples at fixed mass and redshift.

\begin{figure}
\begin{center}
	\includegraphics[width=0.75\linewidth]{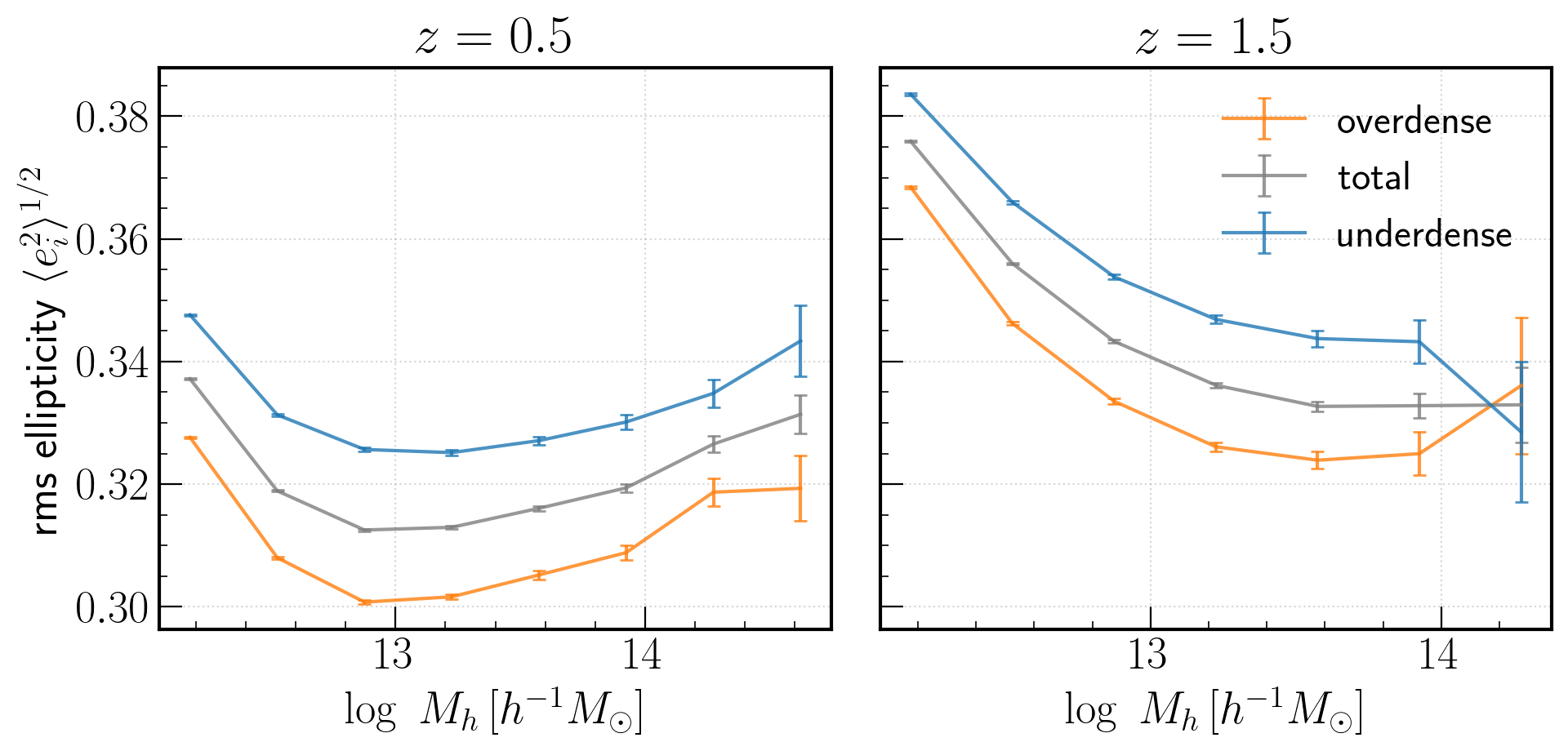}
    \caption{Per-component rms ellipticity $\langle e_i^2 \rangle^{1/2}$ as a function of halo mass for the overdense (orange), total (grey), and underdense (blue) subsamples, at $z=0.5$ (left) and $z=1.5$ (right). At every mass haloes in underdense environments are intrinsically more elongated than haloes of the same mass in overdense environments, showing that the environmental IA signal has a contribution from halo shape and not only from alignment. Error bars are the standard error on the mean per mass bin.}
    \label{fig:erms}
\end{center}
\end{figure}

The per-component rms ellipticity $\langle e_i^2 \rangle^{1/2}$ of the over- and underdense subsamples is listed in Table~\ref{tab:aia} (last three columns) and shown as a function of halo mass in Figure~\ref{fig:erms}. In every mass bin and at both redshifts, haloes in underdense environments are intrinsically more elongated than haloes of the same mass in overdense environments, by between roughly $5$ and $8$ per cent in rms ellipticity. The shape difference grows towards lower redshift and higher halo mass. This demonstrates that the environmental dependence of the IA signal is not driven by alignment alone: a non-negligible part of it arises simply because underdense haloes are less spherical. The same trend is seen in three dimensions: as shown in Figure~\ref{fig:ca} (Appendix~\ref{app:shapes}), at fixed mass the underdense subsample has a systematically smaller minor-to-major axis ratio $c/a$ than the overdense subsample, confirming that haloes in low-density environments are intrinsically more triaxial.

\begin{figure}
\begin{center}
	\includegraphics[width=0.8\linewidth]{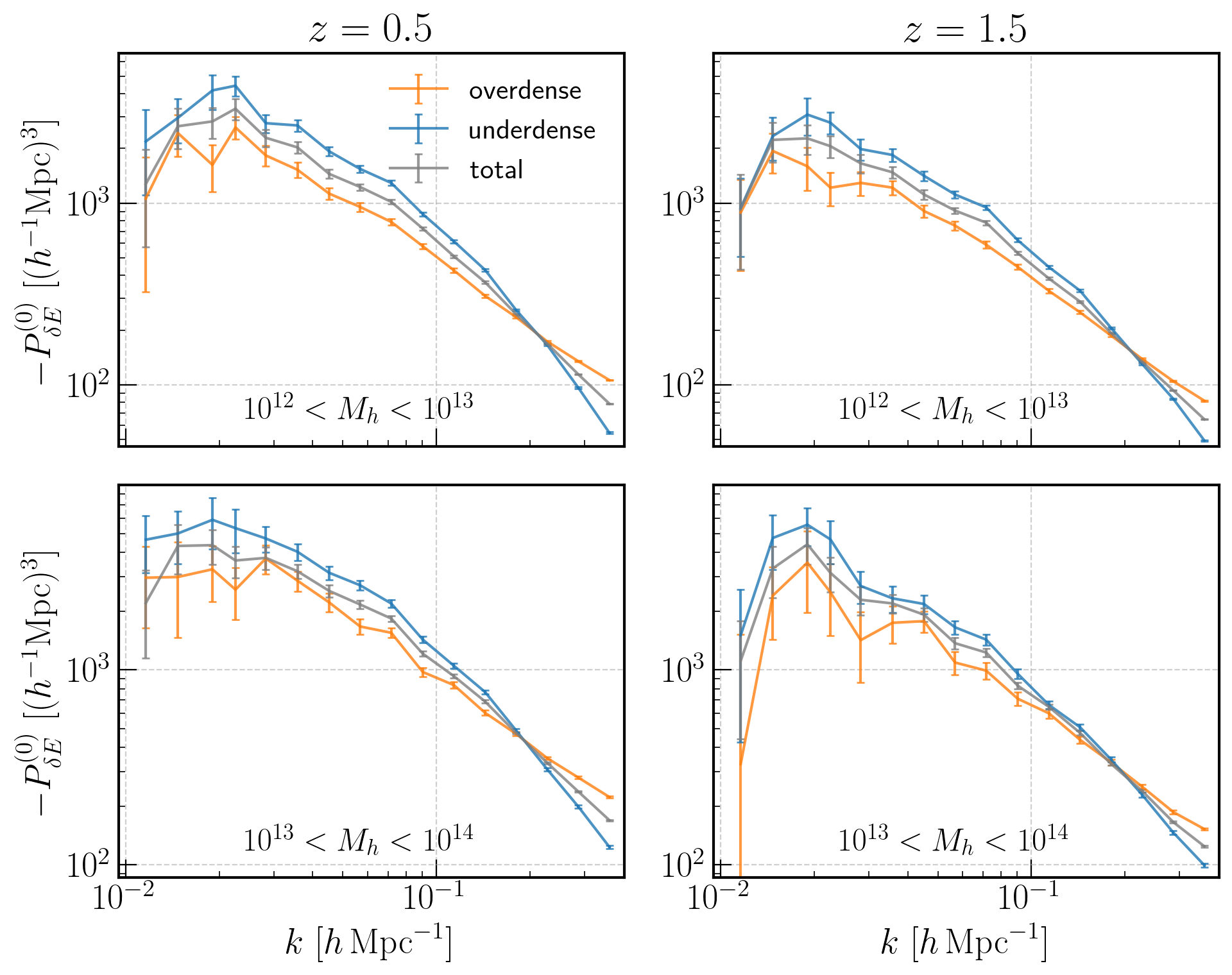}
    \caption{As Figure~\ref{fig:P_de}, but for the orientation-only (unit-ellipticity) shape field: the cross-power spectrum $P^{(0)}_{\delta\hat{E}}$ for the overdense (orange), underdense (blue), and total (grey) subsamples. The environmental ordering persists once the elongation magnitude is removed, demonstrating a genuine difference in halo alignment.}
    \label{fig:P_de_norm}
\end{center}
\end{figure}

To isolate the alignment contribution we repeat the measurement with an orientation-only estimator, in which each halo shape is normalised to unit ellipticity, $\hat{\boldsymbol{e}} = \boldsymbol{e}/|\boldsymbol{e}|$, before its correlation with the density and tidal fields is computed (haloes that are too nearly round for their orientation to be defined are excluded). This removes the dependence on the elongation magnitude, so any remaining environmental difference is due to orientation alignment alone. The resulting normalised cross-power spectrum is shown in Figure~\ref{fig:P_de_norm}.

Fitting these normalised spectra in the same way yields orientation-only amplitude ratios $\hat{A}_{\rm IA}^{\rm under}/\hat{A}_{\rm IA}^{\rm over} \simeq 1.4$--$1.7$, compared with $\simeq 1.5$--$1.8$ for the standard estimator (Table~\ref{tab:aia}). The alignment-only ratio thus accounts for most of the environmental signal, while the residual factor, $\sim5$--$9$ per cent, matches the excess rms ellipticity of the underdense haloes (Table~\ref{tab:aia}). The two effects therefore act in the same direction and combine multiplicatively: haloes in underdense environments are both more strongly aligned with the large-scale tidal field and intrinsically more elongated, with alignment being the dominant of the two.

%%%%%%%%%%%%%%%%%%%%%%%%%%
%Summary and Discussion
%%%%%%%%%%%%%%%%%%%%%%%%%%
\section{Summary and Discussion}
\label{sec:summary}

We have used the high-resolution $N$-body simulations of the \textsc{Dark Quest} suite to measure how the intrinsic alignment of dark matter haloes depends on their large-scale environment over $z=0.1$--$1.5$, characterising the environment by the overdensity $\delta_8$ within $8\,\mpc$ and comparing the most overdense and most underdense haloes within narrow mass bins so that the two subsamples share the same mass distribution by construction. Our main result is that, at fixed halo mass, haloes in underdense environments have systematically larger IA amplitudes than haloes in overdense environments. The underdense-to-overdense ratio of $A_{\rm IA}$ is $\sim1.5$--$1.8$, it is detected in both $P_{\delta E}$ and $P_{EE}$, and it strengthens towards low redshift, growing from $\sim1.6$ at $z=1.5$ to $\sim1.9$ at $z=0.1$ for the lower-mass haloes (at the $\sim8$--$10\sigma$ level). An orientation-only estimator shows that this signal has two origins acting in the same direction: underdense haloes are both more strongly aligned with the large-scale tidal field (orientation-only ratio $\sim1.4$--$1.7$) and intrinsically more elongated ($\sim5$--$8$ per cent larger rms ellipticity, and correspondingly smaller three-dimensional axis ratio $c/a$), with the alignment difference the dominant of the two.

This dependence of IA on environment at fixed halo mass can be regarded as a new manifestation of assembly 
bias~\citep{Sheth_Tormen2004, Gao_2005, Gao_2007, Jing_2007, Dalal_2008, Faltenbacher_White_2010, Shi_Sheth_2018}. Just as the clustering amplitude of haloes of a given mass depends on secondary properties such as formation time, concentration, shape, etc., we find that their intrinsic alignment carries an analogous secondary dependence. Such an ``alignment assembly bias'' is closely related to the shape assembly bias identified by Akitsu et al.~\cite{Akitsu_2021}, who showed that the response of halo shapes to a large-scale tide depends on assembly history; our measurement demonstrates that this effect is large enough to leave a clear, directly measurable imprint on the IA power spectrum itself. We illustrate this explicitly in Appendix~\ref{app:bias} (Fig.~\ref{fig:aia_bias}), where at fixed mass $A_{\rm IA}$ is seen to anti-correlate with the halo bias $b_h$ as the environmental density increases, opposite to the mass sequence along which the two grow together.

At first sight a stronger alignment in underdense regions is counter-intuitive, since the tidal field is weaker there. A plausible interpretation is that the IA amplitude reflects not the instantaneous strength of the tide but how faithfully a halo records the coherent, large-scale tidal field. Haloes in dense, dynamically active environments undergo frequent mergers and experience a rapidly evolving, small-scale-dominated tidal field, which randomises their orientations and erases the large-scale imprint, whereas haloes in quiescent underdense regions retain a cleaner memory of the large-scale tidal field and are also intrinsically more triaxial -- both of which enhance their IA signal. This picture is consistent with Shi et al.~\cite{Shi_2015}, who found halo spins to be more strongly aligned with the tidal field in weaker (more underdense) environments. It is also consistent with the results of Akitsu et al.~\cite{Akitsu_2021}, in which the shape bias is higher for haloes of lower axis ratio (more aspherical; see also Okumura \& Jing~\cite{Teppei_Jing}), and with the finding that more aspherical haloes tend to live in less clustered environments \citep{Faltenbacher_White_2010}. It seems, on the other hand, to be in tension with Xia et al.~\cite{2017ApJ...848...22X}, who reported that haloes in cluster regions are more strongly aligned than those in filaments. However, those authors also showed that dark matter haloes have a lower halo bias (i.e.\ are less clustered) in clusters than in filaments; if we identify haloes of lower halo bias with those in underdense regions, the two results are in fact consistent. The apparent tension arises mainly because the cosmic-web (cluster/filament) classification does not map one-to-one onto the overdensity parameter $\delta_8$ used in this paper. A recent study by Herle et al.~\cite{Herle_2026}, using the \textsc{FLAMINGO} simulation, similarly finds that at fixed halo mass earlier-forming haloes have a larger alignment amplitude. Since such haloes are typically more strongly clustered, this might at first appear to favour stronger alignment in denser environments, opposite to our result; however, their amplitude is measured from the projected position--shape correlation $w_{g+}(r_p)$, which is a density-weighted statistic, so part of the enhancement for the more clustered (older) haloes could arise from the same $(1+\delta_g)$ boost identified above rather than from a genuinely stronger alignment. A like-for-like comparison would require accounting for this boost factor.

These findings have direct implications for the modelling of intrinsic alignments in weak-lensing cosmology. The IA amplitude of a given sample reflects not only the masses and types of its galaxies but also the distribution of environments they occupy, so samples that preferentially select low- or high-density regions may carry systematically different alignment amplitudes even at fixed mass. This is likely to be especially important for beyond-two-point weak-lensing analyses -- such as those based on higher-order shear statistics, density-split or void lensing, and field-level inference -- which deliberately exploit the non-Gaussian, environment-dependent structure of the cosmic web and are therefore more sensitive to an environment-dependent IA contamination than standard two-point analyses. In future work we will quantify directly the impact of this environmental IA dependence on weak-lensing observables, with particular emphasis on beyond-two-point statistics, and investigate whether it can be detected in current and forthcoming observations.

As described above, our results suggest that the IA power spectrum in underdense regions retains more information about the primordial tidal field and therefore contains cleaner cosmological information. This is complementary to the conventional density power spectrum, $P_{\delta\delta}$, 
which is dominated by contributions from overdense regions. 
Therefore, combining the density and IA power spectra, even at the level of two-point statistics, may enable us to extract additional cosmological information
beyond that available from either spectrum alone~\citep[e.g., see][for a related attempt]{2023ApJ...945L..30O}.
Exploring this possibility will be an interesting direction for future work.

\acknowledgments

This work was supported in part by JSPS KAKENHI Grant Number 26K17137, 22K03634, 
24H00215, 24H00221, 26H00401, 26H00402
26H00404, and by World Premier International Research Center Initiative (WPI Initiative), MEXT, Japan. J.S. acknowledges support by the Royal Society through the International Science Partnerships Fund (ISPF) International Collaboration Awards 2023 (Japan) [grant number ICA\textbackslash{}R1\textbackslash{}231094]. J. S. thanks Atsushi Taruya and Teppei Okumura for useful discussions and comments on the draft.

%%%%%%%%%%%%%%%%%%%%%%%%%%%%%%%%%%%%%%%%%%%%%%%%%%
\section*{Data Availability}

The \textsc{Dark Quest} simulation data underlying this article were provided by the \textsc{Dark Quest} collaboration \citep{Nishimichi_2019}. The halo catalogues and the derived shape and IA measurements will be shared on reasonable request to the authors.



\bibliographystyle{JHEP}
\bibliography{ref}

@article{Nishimichi_2019,
   title={Dark Quest. I. Fast and Accurate Emulation of Halo Clustering Statistics and Its Application to Galaxy Clustering},
   volume={884},
   ISSN={1538-4357},
   url={http://dx.doi.org/10.3847/1538-4357/ab3719},
   DOI={10.3847/1538-4357/ab3719},
   number={1},
   journal={The Astrophysical Journal},
   publisher={American Astronomical Society},
   author={Nishimichi, Takahiro and Takada, Masahiro and Takahashi, Ryuichi and Osato, Ken and Shirasaki, Masato and Oogi, Taira and Miyatake, Hironao and Oguri, Masamune and Murata, Ryoma and Kobayashi, Yosuke and Yoshida, Naoki},
   year={2019},
   month=oct, pages={29}
}

@article{ refId0,
	author = {{Planck Collaboration} and {Ade, P. A. R.} and {Aghanim, N.} and {Arnaud, M.} and {Ashdown, M.} and {Aumont, J.} and {Baccigalupi, C.} and {Banday, A. J.} and {Barreiro, R. B.} and {Bartlett, J. G.} and {Bartolo, N.} and {Battaner, E.} and {Battye, R.} and {Benabed, K.} and {Benoît, A.} and {Benoit-Lévy, A.} and {Bernard, J.-P.} and {Bersanelli, M.} and {Bielewicz, P.} and {Bock, J. J.} and {Bonaldi, A.} and {Bonavera, L.} and {Bond, J. R.} and {Borrill, J.} and {Bouchet, F. R.} and {Boulanger, F.} and {Bucher, M.} and {Burigana, C.} and {Butler, R. C.} and {Calabrese, E.} and {Cardoso, J.-F.} and {Catalano, A.} and {Challinor, A.} and {Chamballu, A.} and {Chary, R.-R.} and {Chiang, H. C.} and {Chluba, J.} and {Christensen, P. R.} and {Church, S.} and {Clements, D. L.} and {Colombi, S.} and {Colombo, L. P. L.} and {Combet, C.} and {Coulais, A.} and {Crill, B. P.} and {Curto, A.} and {Cuttaia, F.} and {Danese, L.} and {Davies, R. D.} and {Davis, R. J.} and {de Bernardis, P.} and {de Rosa, A.} and {de Zotti, G.} and {Delabrouille, J.} and {Désert, F.-X.} and {Di Valentino, E.} and {Dickinson, C.} and {Diego, J. M.} and {Dolag, K.} and {Dole, H.} and {Donzelli, S.} and {Doré, O.} and {Douspis, M.} and {Ducout, A.} and {Dunkley, J.} and {Dupac, X.} and {Efstathiou, G.} and {Elsner, F.} and {Enßlin, T. A.} and {Eriksen, H. K.} and {Farhang, M.} and {Fergusson, J.} and {Finelli, F.} and {Forni, O.} and {Frailis, M.} and {Fraisse, A. A.} and {Franceschi, E.} and {Frejsel, A.} and {Galeotta, S.} and {Galli, S.} and {Ganga, K.} and {Gauthier, C.} and {Gerbino, M.} and {Ghosh, T.} and {Giard, M.} and {Giraud-Héraud, Y.} and {Giusarma, E.} and {Gjerløw, E.} and {González-Nuevo, J.} and {Górski, K. M.} and {Gratton, S.} and {Gregorio, A.} and {Gruppuso, A.} and {Gudmundsson, J. E.} and {Hamann, J.} and {Hansen, F. K.} and {Hanson, D.} and {Harrison, D. L.} and {Helou, G.} and {Henrot-Versillé, S.} and {Hernández-Monteagudo, C.} and {Herranz, D.} and {Hildebrandt, S. R.} and {Hivon, E.} and {Hobson, M.} and {Holmes, W. A.} and {Hornstrup, A.} and {Hovest, W.} and {Huang, Z.} and {Huffenberger, K. M.} and {Hurier, G.} and {Jaffe, A. H.} and {Jaffe, T. R.} and {Jones, W. C.} and {Juvela, M.} and {Keihänen, E.} and {Keskitalo, R.} and {Kisner, T. S.} and {Kneissl, R.} and {Knoche, J.} and {Knox, L.} and {Kunz, M.} and {Kurki-Suonio, H.} and {Lagache, G.} and {Lähteenmäki, A.} and {Lamarre, J.-M.} and {Lasenby, A.} and {Lattanzi, M.} and {Lawrence, C. R.} and {Leahy, J. P.} and {Leonardi, R.} and {Lesgourgues, J.} and {Levrier, F.} and {Lewis, A.} and {Liguori, M.} and {Lilje, P. B.} and {Linden-Vørnle, M.} and {López-Caniego, M.} and {Lubin, P. M.} and {Macías-Pérez, J. F.} and {Maggio, G.} and {Maino, D.} and {Mandolesi, N.} and {Mangilli, A.} and {Marchini, A.} and {Maris, M.} and {Martin, P. G.} and {Martinelli, M.} and {Martínez-González, E.} and {Masi, S.} and {Matarrese, S.} and {McGehee, P.} and {Meinhold, P. R.} and {Melchiorri, A.} and {Melin, J.-B.} and {Mendes, L.} and {Mennella, A.} and {Migliaccio, M.} and {Millea, M.} and {Mitra, S.} and {Miville-Deschênes, M.-A.} and {Moneti, A.} and {Montier, L.} and {Morgante, G.} and {Mortlock, D.} and {Moss, A.} and {Munshi, D.} and {Murphy, J. A.} and {Naselsky, P.} and {Nati, F.} and {Natoli, P.} and {Netterfield, C. B.} and {Nørgaard-Nielsen, H. U.} and {Noviello, F.} and {Novikov, D.} and {Novikov, I.} and {Oxborrow, C. A.} and {Paci, F.} and {Pagano, L.} and {Pajot, F.} and {Paladini, R.} and {Paoletti, D.} and {Partridge, B.} and {Pasian, F.} and {Patanchon, G.} and {Pearson, T. J.} and {Perdereau, O.} and {Perotto, L.} and {Perrotta, F.} and {Pettorino, V.} and {Piacentini, F.} and {Piat, M.} and {Pierpaoli, E.} and {Pietrobon, D.} and {Plaszczynski, S.} and {Pointecouteau, E.} and {Polenta, G.} and {Popa, L.} and {Pratt, G. W.} and {Prézeau, G.} and {Prunet, S.} and {Puget, J.-L.} and {Rachen, J. P.} and {Reach, W. T.} and {Rebolo, R.} and {Reinecke, M.} and {Remazeilles, M.} and {Renault, C.} and {Renzi, A.} and {Ristorcelli, I.} and {Rocha, G.} and {Rosset, C.} and {Rossetti, M.} and {Roudier, G.} and {Rouillé d’Orfeuil, B.} and {Rowan-Robinson, M.} and {Rubiño-Martín, J. A.} and {Rusholme, B.} and {Said, N.} and {Salvatelli, V.} and {Salvati, L.} and {Sandri, M.} and {Santos, D.} and {Savelainen, M.} and {Savini, G.} and {Scott, D.} and {Seiffert, M. D.} and {Serra, P.} and {Shellard, E. P. S.} and {Spencer, L. D.} and {Spinelli, M.} and {Stolyarov, V.} and {Stompor, R.} and {Sudiwala, R.} and {Sunyaev, R.} and {Sutton, D.} and {Suur-Uski, A.-S.} and {Sygnet, J.-F.} and {Tauber, J. A.} and {Terenzi, L.} and {Toffolatti, L.} and {Tomasi, M.} and {Tristram, M.} and {Trombetti, T.} and {Tucci, M.} and {Tuovinen, J.} and {Türler, M.} and {Umana, G.} and {Valenziano, L.} and {Valiviita, J.} and {Van Tent, F.} and {Vielva, P.} and {Villa, F.} and {Wade, L. A.} and {Wandelt, B. D.} and {Wehus, I. K.} and {White, M.} and {White, S. D. M.} and {Wilkinson, A.} and {Yvon, D.} and {Zacchei, A.} and {Zonca, A.}},
	title = {Planck 2015 results - XIII. Cosmological parameters},
	DOI= "10.1051/0004-6361/201525830",
	url= "https://doi.org/10.1051/0004-6361/201525830",
	journal = {A\&A},
	year = 2016,
	volume = 594,
	pages = "A13",
}

@article{Lewis_2000,
    doi = {10.1086/309179},
    url = {https://doi.org/10.1086/309179},
    year = {2000},
    month = {aug},
    publisher = {},
    volume = {538},
    number = {2},
    pages = {473},
    author = {Lewis, Antony and Challinor, Anthony and Lasenby, Anthony},
    title = {Efficient Computation of Cosmic Microwave Background Anisotropies in
    Closed Friedmann-Robertson-Walker Models},
    journal = {The Astrophysical Journal}
}

@article{10.1111/j.1365-2966.2006.11040.x,
    author = {Crocce, Martín and Pueblas, Sebastián and Scoccimarro, Román},
    title = {Transients from initial conditions in cosmological simulations},
    journal = {Monthly Notices of the Royal Astronomical Society},
    volume = {373},
    number = {1},
    pages = {369-381},
    year = {2006},
    month = {10},
    issn = {0035-8711},
    doi = {10.1111/j.1365-2966.2006.11040.x},
    url = {https://doi.org/10.1111/j.1365-2966.2006.11040.x},
    eprint = {https://academic.oup.com/mnras/article-pdf/373/1/369/4187617/mnras0373-0369.pdf},
}

@article{10.1046/j.1365-8711.1998.01845.x,
    author = {Scoccimarro, Román},
    title = {Transients from initial conditions: a perturbative analysis},
    journal = {Monthly Notices of the Royal Astronomical Society},
    volume = {299},
    number = {4},
    pages = {1097-1118},
    year = {1998},
    month = {10},
    issn = {0035-8711},
    doi = {10.1046/j.1365-8711.1998.01845.x},
    url = {https://doi.org/10.1046/j.1365-8711.1998.01845.x},
    eprint = {https://academic.oup.com/mnras/article-pdf/299/4/1097/3869550/299-4-1097.pdf},
}

@article{10.1111/j.1365-2966.2005.09655.x,
    author = {Springel, Volker},
    title = {The cosmological simulation code gadget-2},
    journal = {Monthly Notices of the Royal Astronomical Society},
    volume = {364},
    number = {4},
    pages = {1105-1134},
    year = {2005},
    month = {12},
    issn = {0035-8711},
    doi = {10.1111/j.1365-2966.2005.09655.x},
    url = {https://doi.org/10.1111/j.1365-2966.2005.09655.x},
    eprint = {https://academic.oup.com/mnras/article-pdf/364/4/1105/18657201/364-4-1105.pdf},
}

@article{Behroozi_2013,
    doi = {10.1088/0004-637X/762/2/109},
    url = {https://doi.org/10.1088/0004-637X/762/2/109},
    year = {2012},
    month = {dec},
    publisher = {The American Astronomical Society},
    volume = {762},
    number = {2},
    pages = {109},
    author = {Behroozi, Peter S. and Wechsler, Risa H. and Wu, Hao-Yi},
    title = {THE ROCKSTAR PHASE-SPACE TEMPORAL HALO FINDER AND THE VELOCITY OFFSETS OF CLUSTER CORES},
    journal = {The Astrophysical Journal}
}

@article{Abbas_2007,
   title={Strong clustering of underdense regions and the environmental dependence of clustering from Gaussian initial conditions},
   volume={378},
   ISSN={1365-2966},
   url={http://dx.doi.org/10.1111/j.1365-2966.2007.11806.x},
   DOI={10.1111/j.1365-2966.2007.11806.x},
   number={2},
   journal={Monthly Notices of the Royal Astronomical Society},
   publisher={Oxford University Press (OUP)},
   author={Abbas, U. and Sheth, R. K.},
   year={2007},
   month=jun, pages={641–648}
}

@ARTICLE{2023PhRvD.108h3533K,
       author = {{Kurita}, Toshiki and {Takada}, Masahiro},
        title = "{Constraints on anisotropic primordial non-Gaussianity from intrinsic alignments of SDSS-III BOSS galaxies}",
      journal = {\prd},
         year = 2023,
        month = oct,
       volume = {108},
       number = {8},
          eid = {083533},
        pages = {083533},
          doi = {10.1103/PhysRevD.108.083533},
archivePrefix = {arXiv},
       eprint = {2302.02925},
 primaryClass = {astro-ph.CO},
       adsurl = {https://ui.adsabs.harvard.edu/abs/2023PhRvD.108h3533K}
}

@ARTICLE{2022PhRvD.105l3501K,
       author = {{Kurita}, Toshiki and {Takada}, Masahiro},
        title = "{Analysis method for 3D power spectrum of projected tensor fields with fast estimator and window convolution modeling: An application to intrinsic alignments}",
      journal = {\prd},
         year = 2022,
        month = jun,
       volume = {105},
       number = {12},
          eid = {123501},
        pages = {123501},
          doi = {10.1103/PhysRevD.105.123501},
archivePrefix = {arXiv},
       eprint = {2202.11839},
 primaryClass = {astro-ph.CO},
       adsurl = {https://ui.adsabs.harvard.edu/abs/2022PhRvD.105l3501K}
}

@ARTICLE{2021PhRvD.103h3508A,
       author = {{Akitsu}, Kazuyuki and {Kurita}, Toshiki and {Nishimichi}, Takahiro and {Takada}, Masahiro and {Tanaka}, Satoshi},
        title = "{Imprint of anisotropic primordial non-Gaussianity on halo intrinsic alignments in simulations}",
      journal = {\prd},
         year = 2021,
        month = apr,
       volume = {103},
       number = {8},
          eid = {083508},
        pages = {083508},
          doi = {10.1103/PhysRevD.103.083508},
archivePrefix = {arXiv},
       eprint = {2007.03670},
 primaryClass = {astro-ph.CO},
       adsurl = {https://ui.adsabs.harvard.edu/abs/2021PhRvD.103h3508A}
}

@ARTICLE{2021JCAP...05..061V,
       author = {{Vlah}, Zvonimir and {Chisari}, Nora Elisa and {Schmidt}, Fabian},
        title = "{Galaxy shape statistics in the effective field theory}",
      journal = {\jcap},
         year = 2021,
        month = may,
       volume = {2021},
       number = {5},
          eid = {061},
        pages = {061},
          doi = {10.1088/1475-7516/2021/05/061},
archivePrefix = {arXiv},
       eprint = {2012.04114},
 primaryClass = {astro-ph.CO},
       adsurl = {https://ui.adsabs.harvard.edu/abs/2021JCAP...05..061V}
}

@ARTICLE{2023ApJ...945L..30O,
       author = {{Okumura}, Teppei and {Taruya}, Atsushi},
        title = "{First Constraints on Growth Rate from Redshift-space Ellipticity Correlations of SDSS Galaxies at 0.16 < z < 0.70}",
      journal = {\apjl},
         year = 2023,
        month = mar,
       volume = {945},
       number = {2},
          eid = {L30},
        pages = {L30},
          doi = {10.3847/2041-8213/acbf48},
archivePrefix = {arXiv},
       eprint = {2301.06273},
 primaryClass = {astro-ph.CO},
       adsurl = {https://ui.adsabs.harvard.edu/abs/2023ApJ...945L..30O}
}

@article{Kurita_2020,
   title={Power spectrum of halo intrinsic alignments in simulations},
   volume={501},
   ISSN={1365-2966},
   url={http://dx.doi.org/10.1093/mnras/staa3625},
   DOI={10.1093/mnras/staa3625},
   number={1},
   journal={Monthly Notices of the Royal Astronomical Society},
   publisher={Oxford University Press (OUP)},
   author={Kurita, Toshiki and Takada, Masahiro and Nishimichi, Takahiro and Takahashi, Ryuichi and Osato, Ken and Kobayashi, Yosuke},
   year={2020},
   month=nov, pages={833–852}
}

@article{Catelan_2001,
   title={Intrinsic and extrinsic galaxy alignment},
   volume={320},
   ISSN={1365-2966},
   url={http://dx.doi.org/10.1046/j.1365-8711.2001.04105.x},
   DOI={10.1046/j.1365-8711.2001.04105.x},
   number={1},
   journal={Monthly Notices of the Royal Astronomical Society},
   publisher={Oxford University Press (OUP)},
   author={Catelan, P. and Kamionkowski, M. and Blandford, R. D.},
   year={2001},
   month=jan, pages={L7–L13}
}

@article{PhysRevD.70.063526,
  title = {Intrinsic alignment-lensing interference as a contaminant of cosmic shear},
  author = {Hirata, Christopher M. and Seljak, Uro\ifmmode \check{s}\else \v{s}\fi{}},
  journal = {Phys. Rev. D},
  volume = {70},
  issue = {6},
  pages = {063526},
  numpages = {11},
  year = {2004},
  month = {Sep},
  publisher = {American Physical Society},
  doi = {10.1103/PhysRevD.70.063526},
  url = {https://link.aps.org/doi/10.1103/PhysRevD.70.063526}
}

@ARTICLE{2019JCAP...04..031R,
       author = {{Reischke}, Robert and {Sch{\"a}fer}, Bj{\"o}rn Malte},
        title = "{Environmental dependence of ellipticity correlation functions of intrinsic alignments}",
      journal = {\jcap},
         year = 2019,
        month = apr,
       volume = {2019},
       number = {4},
          eid = {031},
        pages = {031},
          doi = {10.1088/1475-7516/2019/04/031},
archivePrefix = {arXiv},
       eprint = {1812.06918},
 primaryClass = {astro-ph.CO},
       adsurl = {https://ui.adsabs.harvard.edu/abs/2019JCAP...04..031R}
}

@ARTICLE{2022MNRAS.509.1985D,
       author = {{d'Assignies D.}, William and {Chisari}, Nora Elisa and {Hamaus}, Nico and {Singh}, Sukhdeep},
        title = "{Intrinsic alignments of galaxies around cosmic voids}",
      journal = {\mnras},
         year = 2022,
        month = jan,
       volume = {509},
       number = {2},
        pages = {1985-1994},
          doi = {10.1093/mnras/stab2986},
archivePrefix = {arXiv},
       eprint = {2108.03922},
 primaryClass = {astro-ph.CO},
       adsurl = {https://ui.adsabs.harvard.edu/abs/2022MNRAS.509.1985D}
}

@ARTICLE{2019A&A...624A..30J,
       author = {{Johnston}, Harry and {Georgiou}, Christos and {Joachimi}, Benjamin and {Hoekstra}, Henk and {Chisari}, Nora Elisa and {Farrow}, Daniel and {Fortuna}, Maria Cristina and {Heymans}, Catherine and {Joudaki}, Shahab and {Kuijken}, Konrad and {Wright}, Angus},
        title = "{KiDS+GAMA: Intrinsic alignment model constraints for current and future weak lensing cosmology}",
      journal = {\aap},
         year = 2019,
        month = apr,
       volume = {624},
          eid = {A30},
        pages = {A30},
          doi = {10.1051/0004-6361/201834714},
archivePrefix = {arXiv},
       eprint = {1811.09598},
 primaryClass = {astro-ph.CO},
       adsurl = {https://ui.adsabs.harvard.edu/abs/2019A&A...624A..30J}
}

@ARTICLE{2015MNRAS.450.2195S,
       author = {{Singh}, Sukhdeep and {Mandelbaum}, Rachel and {More}, Surhud},
        title = "{Intrinsic alignments of SDSS-III BOSS LOWZ sample galaxies}",
      journal = {\mnras},
         year = 2015,
        month = jun,
       volume = {450},
       number = {2},
        pages = {2195-2216},
          doi = {10.1093/mnras/stv778},
archivePrefix = {arXiv},
       eprint = {1411.1755},
 primaryClass = {astro-ph.CO},
       adsurl = {https://ui.adsabs.harvard.edu/abs/2015MNRAS.450.2195S}
}

@ARTICLE{2015PhR...558....1T,
       author = {{Troxel}, M.~A. and {Ishak}, Mustapha},
        title = "{The intrinsic alignment of galaxies and its impact on weak gravitational lensing in an era of precision cosmology}",
      journal = {\physrep},
         year = 2015,
        month = feb,
       volume = {558},
        pages = {1-59},
          doi = {10.1016/j.physrep.2014.11.001},
archivePrefix = {arXiv},
       eprint = {1407.6990},
 primaryClass = {astro-ph.CO},
       adsurl = {https://ui.adsabs.harvard.edu/abs/2015PhR...558....1T}
}

@ARTICLE{2015SSRv..193....1J,
       author = {{Joachimi}, Benjamin and {Cacciato}, Marcello and {Kitching}, Thomas D. and {Leonard}, Adrienne and {Mandelbaum}, Rachel and {Sch{\"a}fer}, Bj{\"o}rn Malte and {Sif{\'o}n}, Crist{\'o}bal and {Hoekstra}, Henk and {Kiessling}, Alina and {Kirk}, Donnacha and {Rassat}, Anais},
        title = "{Galaxy Alignments: An Overview}",
      journal = {\ssr},
         year = 2015,
        month = nov,
       volume = {193},
       number = {1-4},
        pages = {1-65},
          doi = {10.1007/s11214-015-0177-4},
archivePrefix = {arXiv},
       eprint = {1504.05456},
 primaryClass = {astro-ph.GA},
       adsurl = {https://ui.adsabs.harvard.edu/abs/2015SSRv..193....1J}
}

@ARTICLE{2007NJPh....9..444B,
       author = {{Bridle}, Sarah and {King}, Lindsay},
        title = "{Dark energy constraints from cosmic shear power spectra: impact of intrinsic alignments on photometric redshift requirements}",
      journal = {New Journal of Physics},
         year = 2007,
        month = dec,
       volume = {9},
       number = {12},
          eid = {444},
        pages = {444},
          doi = {10.1088/1367-2630/9/12/444},
archivePrefix = {arXiv},
       eprint = {0705.0166},
 primaryClass = {astro-ph},
       adsurl = {https://ui.adsabs.harvard.edu/abs/2007NJPh....9..444B}
}

@ARTICLE{PhysRevD.100.103506,
       author = {{Blazek}, Jonathan A. and {MacCrann}, Niall and {Troxel}, M.~A. and {Fang}, Xiao},
        title = "{Beyond linear galaxy alignments}",
      journal = {\prd},
         year = 2019,
        month = nov,
       volume = {100},
       number = {10},
          eid = {103506},
        pages = {103506},
          doi = {10.1103/PhysRevD.100.103506},
archivePrefix = {arXiv},
       eprint = {1708.09247},
 primaryClass = {astro-ph.CO},
       adsurl = {https://ui.adsabs.harvard.edu/abs/2019PhRvD.100j3506B}
}

@ARTICLE{2002AJ....123..583B,
       author = {{Bernstein}, Gary M. and {Jarvis}, Mike},
        title = "{Shapes and Shears, Stars and Smears: Optimal Measurements for Weak Lensing}",
      journal = {\aj},
         year = 2002,
        month = feb,
       volume = {123},
       number = {2},
        pages = {583-618},
          doi = {10.1086/338085},
archivePrefix = {arXiv},
       eprint = {astro-ph/0107431},
 primaryClass = {astro-ph},
       adsurl = {https://ui.adsabs.harvard.edu/abs/2002AJ....123..583B}
}

@article{Shi_2021_JCAP,
   author = {Shi, Jingjing and Kurita, Toshiki and Takada, Masahiro and Osato, Ken and Kobayashi, Yosuke and Nishimichi, Takahiro},
   title = {Power spectrum of intrinsic alignments of galaxies in IllustrisTNG},
   journal = {Journal of Cosmology and Astroparticle Physics},
   year = {2021},
   volume = {2021},
   number = {03},
   pages = {030},
   doi = {10.1088/1475-7516/2021/03/030},
   eprint = {2009.00276},
   archivePrefix = {arXiv},
   primaryClass = {astro-ph.CO}
}

@ARTICLE{2017ApJ...848...22X,
       author = {{Xia}, Qianli and {Kang}, Xi and {Wang}, Peng and {Luo}, Yu and {Yang}, Xiaohu and {Jing}, Yipeng and {Wang}, Huiyuan and {Mo}, Houjun},
        title = "{Halo Intrinsic Alignment: Dependence on Mass, Formation Time, and Environment}",
      journal = {\apj},
         year = 2017,
        month = oct,
       volume = {848},
       number = {1},
          eid = {22},
        pages = {22},
          doi = {10.3847/1538-4357/aa8d17},
archivePrefix = {arXiv},
       eprint = {1710.01730},
 primaryClass = {astro-ph.CO},
       adsurl = {https://ui.adsabs.harvard.edu/abs/2017ApJ...848...22X}
}

@ARTICLE{Bartelmann_Schneider_2001,
       author = {{Bartelmann}, Matthias and {Schneider}, Peter},
        title = "{Weak gravitational lensing}",
      journal = {\physrep},
         year = 2001,
       volume = {340},
       number = {4-5},
        pages = {291-472},
          doi = {10.1016/S0370-1573(00)00082-X},
archivePrefix = {arXiv},
       eprint = {astro-ph/9912508},
 primaryClass = {astro-ph}
}

@ARTICLE{Hoekstra_Jain_2008,
       author = {{Hoekstra}, Henk and {Jain}, Bhuvnesh},
        title = "{Weak Gravitational Lensing and its Cosmological Applications}",
      journal = {Annual Review of Nuclear and Particle Science},
         year = 2008,
       volume = {58},
       number = {1},
        pages = {99-123},
          doi = {10.1146/annurev.nucl.58.110707.171151},
archivePrefix = {arXiv},
       eprint = {0805.0139},
 primaryClass = {astro-ph}
}

@ARTICLE{Kilbinger_2015,
       author = {{Kilbinger}, Martin},
        title = "{Cosmology with cosmic shear observations: a review}",
      journal = {Reports on Progress in Physics},
         year = 2015,
       volume = {78},
       number = {8},
          eid = {086901},
        pages = {086901},
          doi = {10.1088/0034-4885/78/8/086901},
archivePrefix = {arXiv},
       eprint = {1411.0115},
 primaryClass = {astro-ph.CO}
}

@ARTICLE{Teppei_Jing,
       author = {{Okumura}, Teppei and {Jing}, Y.~P.},
        title = "{The Gravitational Shear-Intrinsic Ellipticity Correlation Functions of Luminous Red Galaxies in Observation and in the {$\Lambda$}CDM Model}",
      journal = {\apjl},
         year = 2009,
        month = mar,
       volume = {694},
       number = {1},
        pages = {L83-L86},
          doi = {10.1088/0004-637X/694/1/L83},
archivePrefix = {arXiv},
       eprint = {0812.2935},
 primaryClass = {astro-ph}
}

@ARTICLE{Krause_2016,
       author = {{Krause}, Elisabeth and {Eifler}, Tim and {Blazek}, Jonathan},
        title = "{The impact of intrinsic alignment on current and future cosmic shear surveys}",
      journal = {\mnras},
         year = 2016,
       volume = {456},
       number = {1},
        pages = {207-222},
          doi = {10.1093/mnras/stv2615},
archivePrefix = {arXiv},
       eprint = {1506.08730},
 primaryClass = {astro-ph.CO}
}

@ARTICLE{VanAlfen_2024,
       author = {{Van Alfen}, Nicholas and {Campbell}, Duncan and {Blazek}, Jonathan and {Leonard}, C.~Danielle and {Lanusse}, Fran\c{c}ois and {Hearin}, Andrew and {Mandelbaum}, Rachel},
        title = "{An Empirical Model For Intrinsic Alignments: Insights From Cosmological Simulations}",
      journal = {The Open Journal of Astrophysics},
         year = 2024,
       volume = {7},
          doi = {10.33232/001c.118783},
archivePrefix = {arXiv},
       eprint = {2311.07374},
 primaryClass = {astro-ph.CO}
}

@ARTICLE{Herle_2026,
       author = {{Herle}, A. and {Chisari}, N.~E. and {Hoekstra}, H. and {Neumann}, D.},
        title = "{Assembly bias and the redshift evolution of intrinsic alignments for LRGs}",
      journal = {arXiv e-prints},
         year = 2026,
          eid = {arXiv:2607.00785},
        pages = {arXiv:2607.00785},
archivePrefix = {arXiv:2607.00785},
       eprint = {2607.00785},
 primaryClass = {astro-ph.CO},
 adsurl = {https://ui.adsabs.harvard.edu/abs/2026arXiv260700785H/abstract}
}

@ARTICLE{Lamman_2024,
       author = {{Lamman}, Claire and {Tsaprazi}, Eleni and {Shi}, Jingjing and {{\v S}ar{\v c}evi{\'c}}, Nikolina Niko and {Pyne}, Susan and {Legnani}, Elisa and {Ferreira}, Tassia},
        title = "{The IA Guide: A Breakdown of Intrinsic Alignment Formalisms}",
      journal = {The Open Journal of Astrophysics},
         year = 2024,
       volume = {7},
          doi = {10.21105/astro.2309.08605},
archivePrefix = {arXiv},
       eprint = {2309.08605},
 primaryClass = {astro-ph.CO}
}

@ARTICLE{Akitsu_2023,
       author = {{Akitsu}, Kazuyuki and {Li}, Yin and {Okumura}, Teppei},
        title = "{Quadratic shape biases in three-dimensional halo intrinsic alignments}",
      journal = {\jcap},
         year = 2023,
       volume = {2023},
       number = {08},
        pages = {068},
          doi = {10.1088/1475-7516/2023/08/068},
archivePrefix = {arXiv},
       eprint = {2306.00969},
 primaryClass = {astro-ph.CO}
}

@ARTICLE{Akitsu_2021,
       author = {{Akitsu}, Kazuyuki and {Li}, Yin and {Okumura}, Teppei},
        title = "{Cosmological simulation in tides: power spectra, halo shape responses, and shape assembly bias}",
      journal = {\jcap},
         year = 2021,
       volume = {2021},
       number = {04},
        pages = {041},
          doi = {10.1088/1475-7516/2021/04/041},
archivePrefix = {arXiv},
       eprint = {2011.06584},
 primaryClass = {astro-ph.CO}
}

@ARTICLE{Shi_2015,
       author = {{Shi}, Jingjing and {Wang}, Huiyuan and {Mo}, H.~J.},
        title = "{Flow Patterns around Dark Matter Halos: The Link between Halo Dynamical Properties and Large-scale Tidal Field}",
      journal = {\apj},
         year = 2015,
        month = jul,
       volume = {807},
       number = {1},
          eid = {37},
        pages = {37},
          doi = {10.1088/0004-637X/807/1/37},
archivePrefix = {arXiv},
       eprint = {1501.07764},
 primaryClass = {astro-ph.CO},
       adsurl = {https://ui.adsabs.harvard.edu/abs/2015ApJ...807...37S}
}

@ARTICLE{Sheth_Tormen2004,
       author = {{Sheth}, Ravi K. and {Tormen}, Giuseppe},
        title = "{On the environmental dependence of halo formation}",
      journal = {\mnras},
         year = 2004,
        month = jun,
       volume = {350},
       number = {4},
        pages = {1385-1390},
          doi = {10.1111/j.1365-2966.2004.07733.x},
archivePrefix = {arXiv},
       eprint = {astro-ph/0402237},
 primaryClass = {astro-ph},
       adsurl = {https://ui.adsabs.harvard.edu/abs/2004MNRAS.350.1385S}
}

@ARTICLE{Dalal_2008,
       author = {{Dalal}, Neal and {White}, Martin and {Bond}, J. Richard and {Shirokov}, Alexander},
        title = "{Halo Assembly Bias in Hierarchical Structure Formation}",
      journal = {\apj},
         year = 2008,
        month = nov,
       volume = {687},
       number = {1},
        pages = {12-21},
          doi = {10.1086/591512},
archivePrefix = {arXiv},
       eprint = {0803.3453},
 primaryClass = {astro-ph},
       adsurl = {https://ui.adsabs.harvard.edu/abs/2008ApJ...687...12D}
}

@ARTICLE{Gao_2005,
       author = {{Gao}, Liang and {Springel}, Volker and {White}, Simon D.~M.},
        title = "{The age dependence of halo clustering}",
      journal = {\mnras},
         year = 2005,
        month = oct,
       volume = {363},
       number = {1},
        pages = {L66-L70},
          doi = {10.1111/j.1745-3933.2005.00084.x},
archivePrefix = {arXiv},
       eprint = {astro-ph/0506510},
 primaryClass = {astro-ph},
       adsurl = {https://ui.adsabs.harvard.edu/abs/2005MNRAS.363L..66G}
}

@ARTICLE{Gao_2007,
       author = {{Gao}, Liang and {White}, Simon D.~M.},
        title = "{Assembly bias in the clustering of dark matter haloes}",
      journal = {\mnras},
         year = 2007,
        month = apr,
       volume = {377},
       number = {1},
        pages = {L5-L9},
          doi = {10.1111/j.1745-3933.2007.00292.x},
archivePrefix = {arXiv},
       eprint = {astro-ph/0611921},
 primaryClass = {astro-ph},
       adsurl = {https://ui.adsabs.harvard.edu/abs/2007MNRAS.377L...5G}
}

@ARTICLE{Jing_2007,
       author = {{Jing}, Y.~P. and {Suto}, Yasushi and {Mo}, H.~J.},
        title = "{The Dependence of Dark Halo Clustering on Formation Epoch and Concentration Parameter}",
      journal = {\apj},
         year = 2007,
        month = mar,
       volume = {657},
       number = {2},
        pages = {664-668},
          doi = {10.1086/511130},
archivePrefix = {arXiv},
       eprint = {astro-ph/0610099},
 primaryClass = {astro-ph},
       adsurl = {https://ui.adsabs.harvard.edu/abs/2007ApJ...657..664J}
}

@ARTICLE{Shi_Sheth_2018,
       author = {{Shi}, Jingjing and {Sheth}, Ravi K.},
        title = "{Dependence of halo bias on mass and environment}",
      journal = {\mnras},
         year = 2018,
        month = jan,
       volume = {473},
       number = {2},
        pages = {2486-2492},
          doi = {10.1093/mnras/stx2277},
archivePrefix = {arXiv},
       eprint = {1707.04096},
 primaryClass = {astro-ph.CO},
       adsurl = {https://ui.adsabs.harvard.edu/abs/2018MNRAS.473.2486S}
}

@ARTICLE{Faltenbacher_White_2010,
       author = {{Faltenbacher}, Andreas and {White}, Simon D.~M.},
        title = "{Assembly Bias and the Dynamical Structure of Dark Matter Halos}",
      journal = {\apj},
         year = 2010,
        month = jan,
       volume = {708},
       number = {1},
        pages = {469-473},
          doi = {10.1088/0004-637X/708/1/469},
archivePrefix = {arXiv},
       eprint = {0909.4302},
 primaryClass = {astro-ph.CO},
       adsurl = {https://ui.adsabs.harvard.edu/abs/2010ApJ...708..469F}
}

@ARTICLE{Chisari_2025,
       author = {{Chisari}, Nora Elisa},
        title = "{A rising tide: intrinsic alignments since the turn of the millennium}",
      journal = {\aapr},
         year = 2025,
        month = oct,
       volume = {33},
       number = {1},
          eid = {5},
        pages = {5},
          doi = {10.1007/s00159-025-00161-8},
archivePrefix = {arXiv},
       eprint = {2510.15738},
 primaryClass = {astro-ph.CO},
       adsurl = {https://ui.adsabs.harvard.edu/abs/2025A&ARv..33....5C}
}

@ARTICLE{Vlah+2020:IA_EFT,
       author = {{Vlah}, Zvonimir and {Chisari}, Nora Elisa and {Schmidt}, Fabian},
        title = "{An EFT description of galaxy intrinsic alignments}",
      journal = {\jcap},
         year = 2020,
        month = jan,
       volume = {2020},
       number = {1},
          eid = {025},
        pages = {025},
          doi = {10.1088/1475-7516/2020/01/025},
archivePrefix = {arXiv},
       eprint = {1910.08085},
 primaryClass = {astro-ph.CO},
       adsurl = {https://ui.adsabs.harvard.edu/abs/2020JCAP...01..025V}
}

@article{Vlah+2021:IA_EFT,
   title={Galaxy shape statistics in the effective field theory},
   volume={2021},
   ISSN={1475-7516},
   url={http://dx.doi.org/10.1088/1475-7516/2021/05/061},
   DOI={10.1088/1475-7516/2021/05/061},
   number={05},
   journal={Journal of Cosmology and Astroparticle Physics},
   publisher={IOP Publishing},
   author={Vlah, Zvonimir and Chisari, Nora Elisa and Schmidt, Fabian},
   year={2021},
   month={May},
   pages={061}
}

@ARTICLE{Bakx+2023:EFTofIAvsSims,
       author = {{Bakx}, Thomas and {Kurita}, Toshiki and {Elisa Chisari}, Nora and {Vlah}, Zvonimir and {Schmidt}, Fabian},
        title = "{Effective field theory of intrinsic alignments at one loop order: a comparison to dark matter simulations}",
      journal = {\jcap},
         year = 2023,
        month = oct,
       volume = {2023},
       number = {10},
          eid = {005},
        pages = {005},
          doi = {10.1088/1475-7516/2023/10/005},
archivePrefix = {arXiv},
       eprint = {2303.15565},
 primaryClass = {astro-ph.CO},
       adsurl = {https://ui.adsabs.harvard.edu/abs/2023JCAP...10..005B}
}

@ARTICLE{Chen&Kokron2024:LEFTofIA,
       author = {{Chen}, Shi-Fan and {Kokron}, Nickolas},
        title = "{A Lagrangian theory for galaxy shape statistics}",
      journal = {\jcap},
         year = 2024,
        month = jan,
       volume = {2024},
       number = {1},
          eid = {027},
        pages = {027},
          doi = {10.1088/1475-7516/2024/01/027},
archivePrefix = {arXiv},
       eprint = {2309.16761},
 primaryClass = {astro-ph.CO},
       adsurl = {https://ui.adsabs.harvard.edu/abs/2024JCAP...01..027C}
}

@ARTICLE{Kurita_2026,
       author = {{Kurita}, Toshiki and {Jamieson}, Drew and {Komatsu}, Eiichiro and {Schmidt}, Fabian},
        title = "{Parity violation in galaxy shapes: Primordial non-Gaussianity}",
      journal = {\prd},
         year = 2026,
        month = mar,
       volume = {113},
       number = {6},
          eid = {063557},
        pages = {063557},
          doi = {10.1103/fxh6-hpmk},
archivePrefix = {arXiv},
       eprint = {2509.08787},
 primaryClass = {astro-ph.CO},
       adsurl = {https://ui.adsabs.harvard.edu/abs/2026PhRvD.113f3557K}
}

@ARTICLE{Siegel_2025,
       author = {{Siegel}, J. and {McCullough}, J. and {Amon}, A. and {Lamman}, C. and {Jeffrey}, N. and {Joachimi}, B. and {Hoekstra}, H. and {Heydenreich}, S. and {Ross}, A.~J. and {Aguilar}, J. and {Ahlen}, S. and {Bianchi}, D. and {Blake}, C. and {Brooks}, D. and {Castander}, F.~J. and {Claybaugh}, T. and {de la Macorra}, A. and {DeRose}, J. and {Doel}, P. and {Emas}, N. and {Ferraro}, S. and {Font-Ribera}, A. and {Forero-Romero}, J.~E. and {Gazta{\~n}aga}, E. and {Gontcho}, S. Gontcho A and {Gutierrez}, G. and {Honscheid}, K. and {Ishak}, M. and {Joudaki}, S. and {Kehoe}, R. and {Kirkby}, D. and {Kisner}, T. and {Krolewski}, A. and {Lahav}, O. and {Lambert}, A. and {Landriau}, M. and {Le Guillou}, L. and {Levi}, M.~E. and {Manera}, M. and {Meisner}, A. and {Miquel}, R. and {Moustakas}, J. and {Nadathur}, S. and {Newman}, J.~A. and {Niz}, G. and {Palanque-Delabrouille}, N. and {Percival}, W.~J. and {Porredon}, A. and {Prada}, F. and {P{\'e}rez-R{\`a}fols}, I. and {Rossi}, G. and {Sanchez}, E. and {Saulder}, C. and {Schlegel}, D. and {Schubnell}, M. and {Semenaite}, A. and {Silber}, J. and {Sprayberry}, D. and {Sun}, Z. and {Tarl{\'e}}, G. and {Weaver}, B.~A. and {Zhou}, R. and {Zou}, H.},
        title = "{Intrinsic alignment demographics for next-generation lensing: Revealing galaxy property trends with DESI Y1 direct measurements}",
      journal = {arXiv e-prints},
         year = 2025,
        month = jul,
          eid = {arXiv:2507.11530},
        pages = {arXiv:2507.11530},
          doi = {10.48550/arXiv.2507.11530},
archivePrefix = {arXiv},
       eprint = {2507.11530},
 primaryClass = {astro-ph.CO},
       adsurl = {https://ui.adsabs.harvard.edu/abs/2025arXiv250711530S}
}

@ARTICLE{Osato_2018,
       author = {{Osato}, Ken and {Nishimichi}, Takahiro and {Oguri}, Masamune and {Takada}, Masahiro and {Okumura}, Teppei},
        title = "{Strong orientation dependence of surface mass density profiles of dark haloes at large scales}",
      journal = {\mnras},
         year = 2018,
        month = jun,
       volume = {477},
       number = {2},
        pages = {2141-2153},
          doi = {10.1093/mnras/sty762},
archivePrefix = {arXiv},
       eprint = {1712.00094},
 primaryClass = {astro-ph.CO},
       adsurl = {https://ui.adsabs.harvard.edu/abs/2018MNRAS.477.2141O}
}

%\begin{thebibliography}{99}
%\end{thebibliography}

%%%%%%%%%%%%%%%%%%%%%%%%%%%%%%%%%%%%%%%%%%%%%%%%%%

%%%%%%%%%%%%%%%%% APPENDICES %%%%%%%%%%%%%%%%%%%%%

\appendix

\section{Non-Poisson noise in $P_{EE}$}
\label{app:extra}

The shape auto-power spectra $P_{EE}$ and $P_{BB}$ both contain a shape-noise contribution that arises from the finite number of haloes sampling the shape field \citep[see][their Appendix~B]{Kurita_2020}. In the linear-alignment model the intrinsic shape responds to a 
scalar tidal field and is therefore curl-free, so the $B$-mode carries no IA signal on large scales: $\langle\gamma_B\gamma_B\rangle=0$ in the linear regime, while the cross-spectra $\langle\gamma_B\delta\rangle$ and $\langle\gamma_E\gamma_B\rangle$ vanish by statistical parity invariance. On large (linear) scales the measured $P_{BB}$ is thus signal-free and provides an independent, in-situ estimate of the noise in $P_{EE}$. On smaller, non-linear scales this no longer holds exactly, as non-linear evolution of the alignment can source a genuine $B$-mode and $P_{BB}$ need not vanish; we therefore use $P_{BB}$ as a noise estimator only on the large scales relevant to our fit.

\begin{figure}
\begin{center}
	\includegraphics[width=0.6\columnwidth]{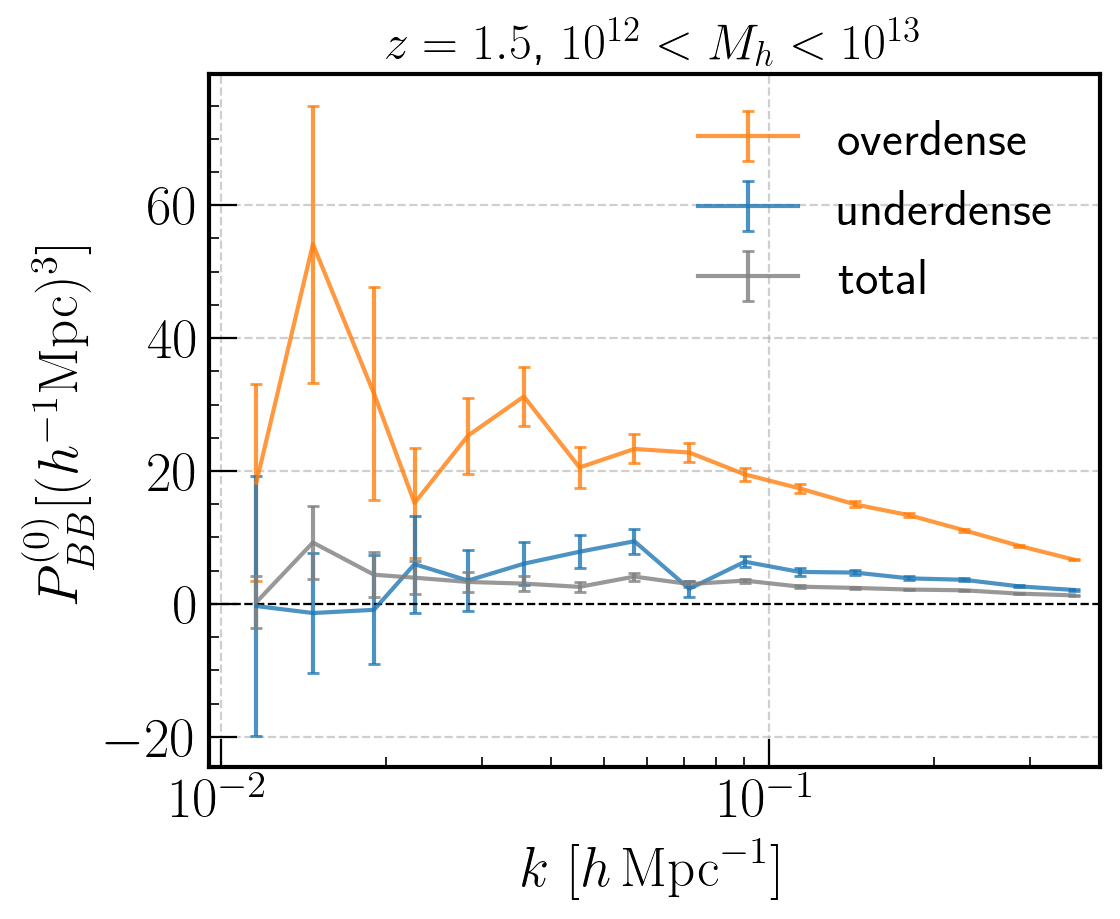}
    \caption{Residual $B$-mode auto-power spectrum after subtracting the analytic Poisson shot noise, $\langle \sigma_\gamma^2 \rangle / \bar n_h$, for the overdense (orange), underdense (blue), and total (grey) subsamples in the $10^{12} < M_h < 10^{13}\,\msun$ bin at $z=1.5$. Since the signal is $B$-free, a non-zero residual measures non-Poisson noise at small $k$: it is consistent with zero for the underdense and total subsamples but significantly positive for the overdense subsample, showing that the excess noise in $P_{EE}$ originates predominantly from the strongly clustered overdense haloes.}
    \label{fig:pbb_resid}
\end{center}
\end{figure}

This noise has two parts \citep[][their Appendix~B]{Kurita_2020}: the standard Poisson shot noise, $P_{\rm shot} = \langle \sigma_\gamma^2 \rangle / \bar n_h$, that would arise if halo shapes were mutually uncorrelated, plus a \emph{non-Poisson} contribution that reflects the clustering of the haloes whose shapes are sampled. Kurita et al.~\cite{Kurita_2020} found that the measured $B$-mode exceeds the Poisson prediction by roughly $5$--$10$ per cent, a ``renormalised'' offset arising from the $k\to0$ limit of higher-order contributions to the $B$-mode spectrum. Any departure of $P_{BB}$ from $P_{\rm shot}$ therefore measures this non-Poisson noise. Figure~\ref{fig:pbb_resid} shows the residual, $(P_{BB} - P_{\rm shot})$, for the $10^{12} < M_h < 10^{13}\,\msun$ bin at $z=1.5$. For the underdense and total subsamples the residual is consistent with zero at large scales, i.e.\ the noise is essentially Poisson; for the overdense subsample, however, it is significantly positive. Overdense haloes are strongly clustered, so their shape sampling is far from a uniform Poisson process; this excess noise is largest for the overdense sample and, because massive high-redshift bins are sparse, it is most visible in the $z=1.5$, high-mass panel of Figure~\ref{fig:P_ee}.

\begin{figure}
\begin{center}
	\includegraphics[width=0.8\linewidth]{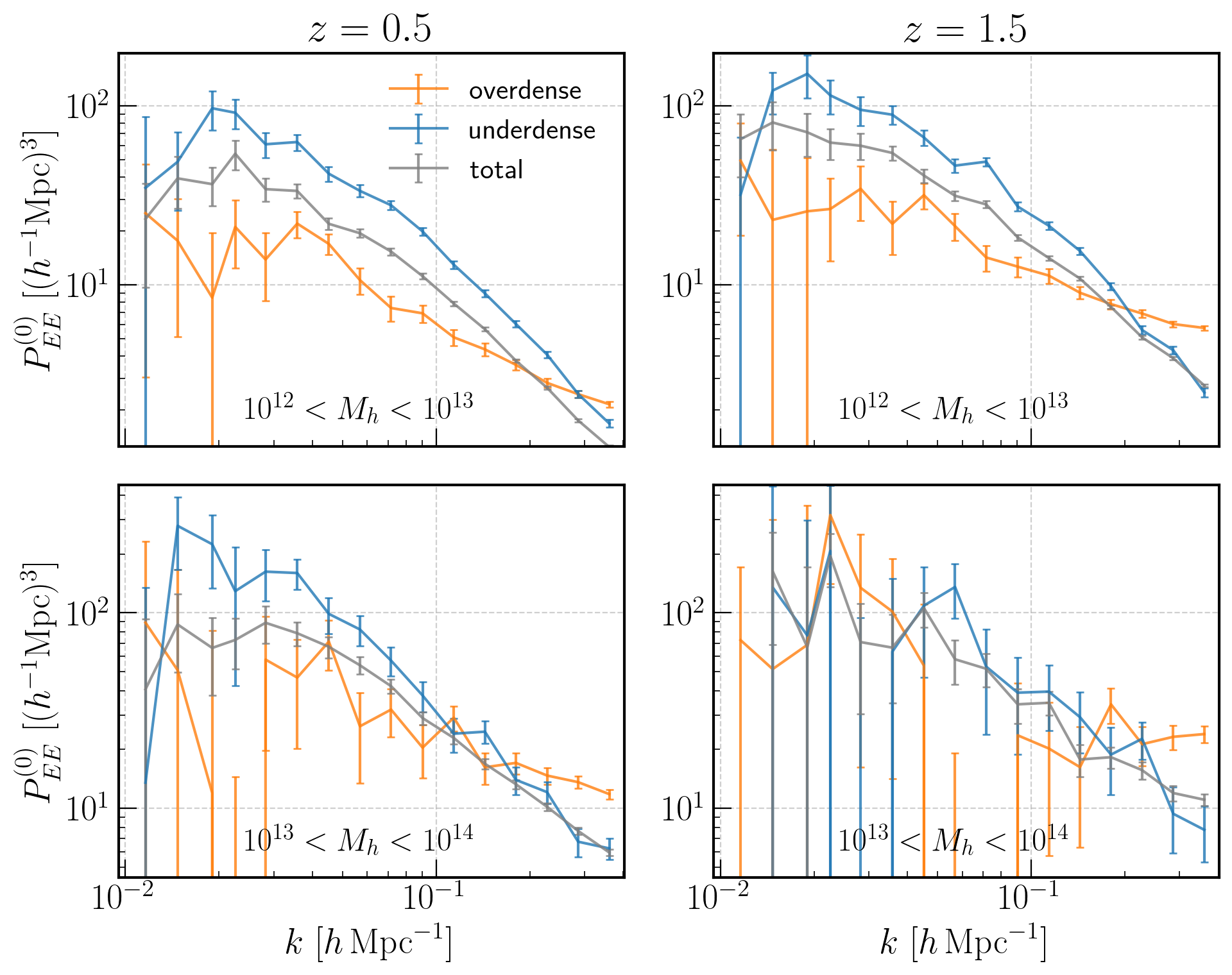}
    \caption{As Figure~\ref{fig:P_ee}, but with the noise removed by subtracting the measured $B$-mode auto-power spectrum, $P_{EE} - P_{BB}$, instead of the analytic Poisson term. This additionally removes the non-Poisson noise (Figure~\ref{fig:pbb_resid}), so the spurious high-noise behaviour of the overdense sample is suppressed; the error bars are correspondingly larger because the $B$-mode variance is added in quadrature.}
    \label{fig:P_ee_bmode}
\end{center}
\end{figure}

Since shape noise is statistically isotropic between the $E$ and $B$ modes, this non-Poisson component can be removed empirically by subtracting the measured $B$-mode rather than the analytic Poisson term, i.e.\ $P_{EE} - P_{BB}$ in place of $P_{EE} - P_{\rm shot}$. However, we should point out that the small scale suffers from oversubstraction since the small scale $BB$ signal is mostly physical instead of from pure noise. Since we are mostly focused on large scale in this work, we will ignore such oversubstraction for now. Figure~\ref{fig:P_ee_bmode} shows the result: the residual noise (and the spurious excess of the overdense curve) is suppressed relative to Figure~\ref{fig:P_ee}, at the cost of larger error bars, since the $B$-mode variance adds in quadrature. The large-scale environmental ordering, underdense $>$ overdense, is unchanged. We note that the $B$-mode subtraction removes only the noise; the overdense $P_{EE}$ retains the $(1+\delta_h)^2$ density-weighting boost of the signal itself (Section~\ref{sec:results_main}), which inflates it by a further $\sim40$ per cent at $k\simeq0.02\,\hmpci$ (and more towards smaller scales) while leaving the far less biased underdense sample almost unaffected. The environmental contrast in Figure~\ref{fig:P_ee_bmode} should therefore be regarded as a lower bound on the true, boost-free difference.

\section{Dependence on the shape estimator}
\label{app:estimator}

\begin{figure}
\begin{center}
	\includegraphics[width=0.8\linewidth]{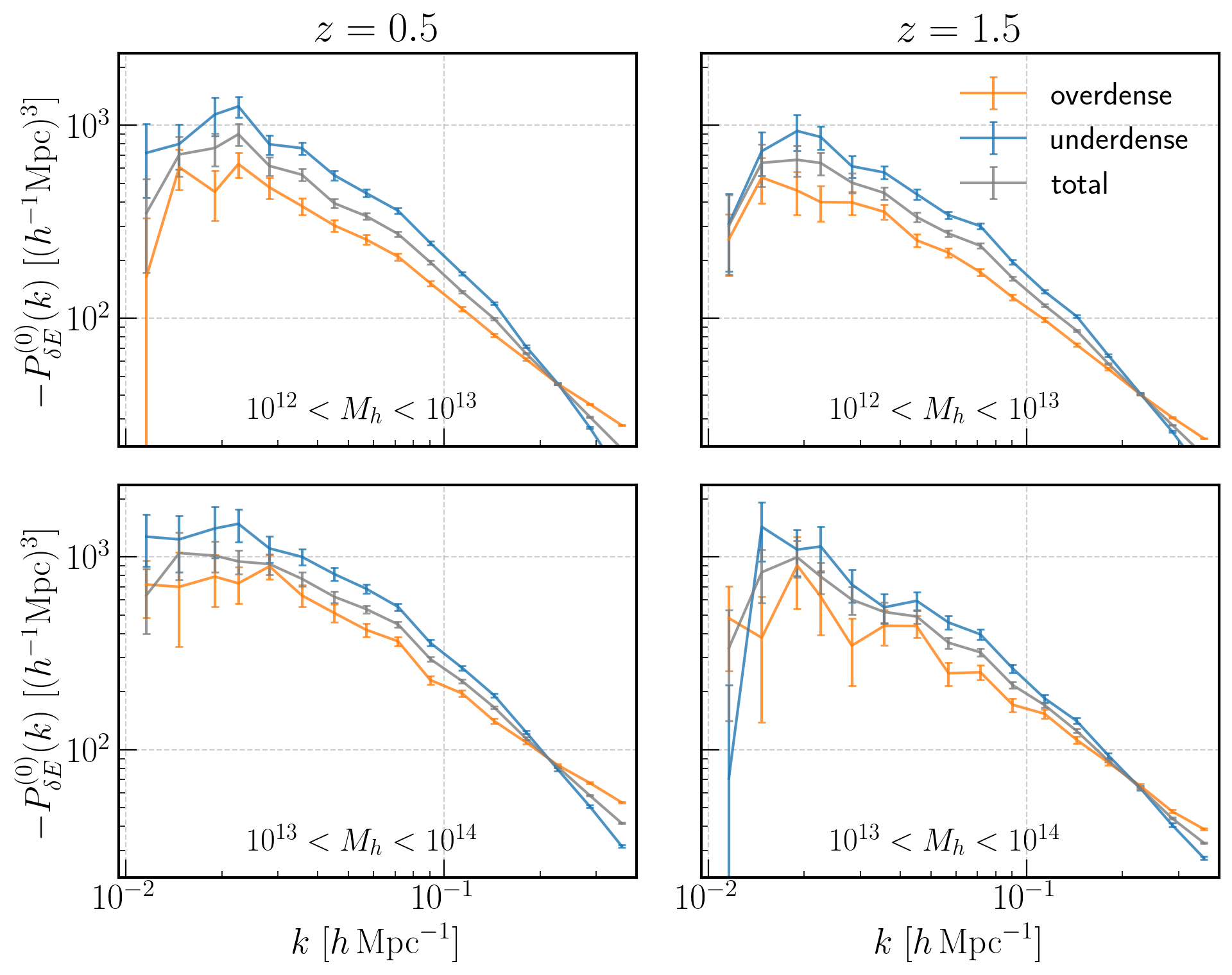}
    \caption{As Figure~\ref{fig:P_de} ($-P^{(0)}_{\delta E}$ for the overdense (orange), underdense (blue), and total (grey) subsamples, with statistical error bars), but with halo shapes measured from the unreduced (simple) iterative inertia tensor rather than the fiducial reduced inertia tensor. The environmental ordering matches Figure~\ref{fig:P_de}, confirming that the result is insensitive to the shape estimator.}
    \label{fig:P_de_SIAC}
\end{center}
\end{figure}

Our fiducial halo shapes are measured from the reduced inertia tensor computed with the iterative method (Section~\ref{sec:data}, equation~\ref{eq:inertia}). As shown by Kurita et al.~\cite[their Appendix~C]{Kurita_2020}, different inertia-tensor definitions differ mainly by an overall amplitude while sharing the same $k$-dependence on large scales. To confirm that our environmental result is not an artefact of this choice, we repeat the $P_{\delta E}$ measurement using an alternative estimator: the \emph{unreduced} (simple) inertia tensor, again computed iteratively -- i.e.\ equation~(\ref{eq:inertia}) without the $1/r_p^2$ radial weighting. Figure~\ref{fig:P_de_SIAC} shows the result for the same mass ranges and redshifts as Figure~\ref{fig:P_de}. The environmental ordering is unchanged: on large scales the underdense subsample has the strongest signal and the overdense the weakest, with the total in between, and the same small-scale reversal appears. The environmental dependence reported in Section~\ref{sec:results} is therefore robust to the choice of shape estimator.

\section{Three-dimensional halo shapes}
\label{app:shapes}

\begin{figure}
\begin{center}
	\includegraphics[width=0.75\linewidth]{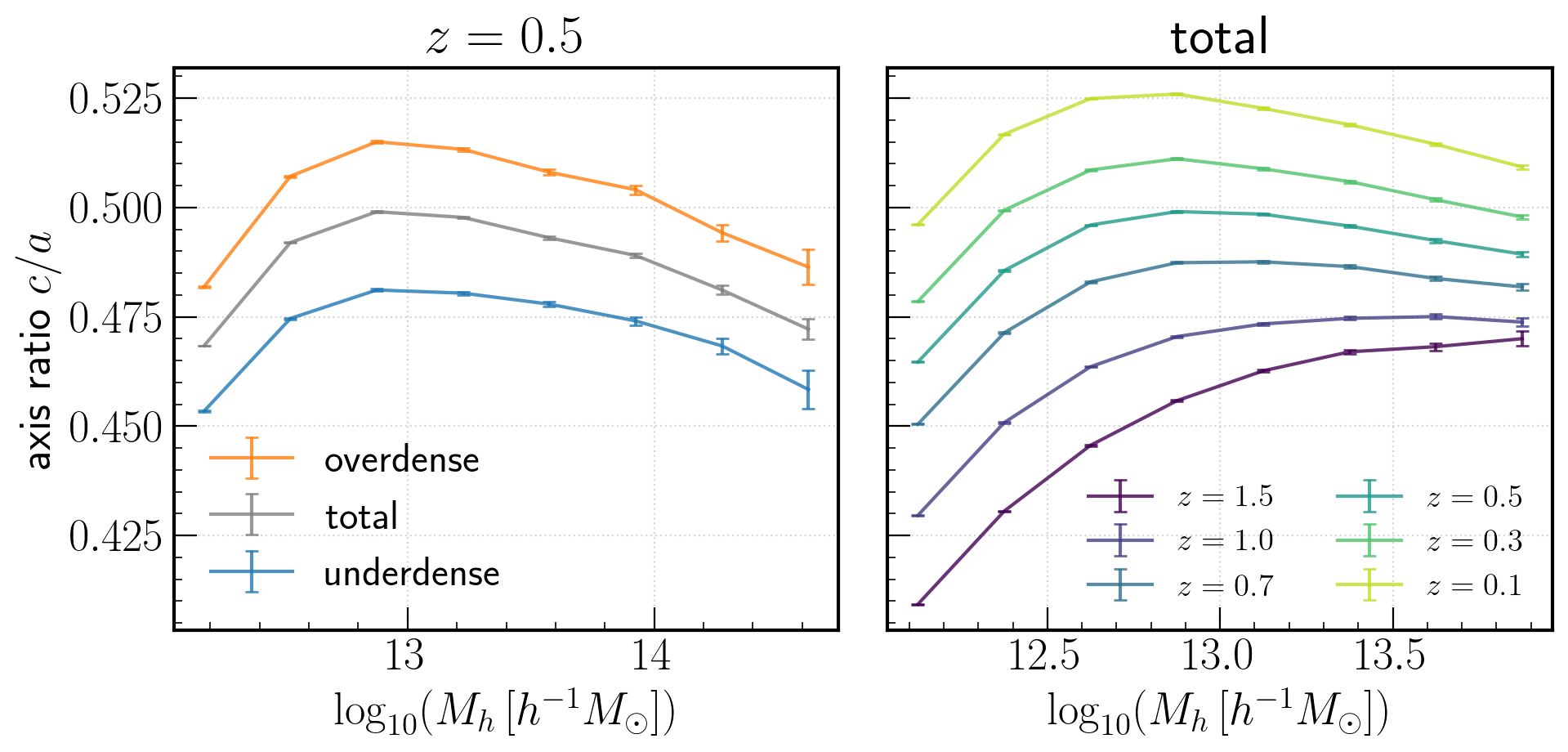}
    \caption{Mean three-dimensional minor-to-major axis ratio $c/a$ per mass bin, from the eigenvalues of the halo inertia tensor. Left: the overdense (orange), total (grey), and underdense (blue) subsamples at $z=0.5$ as a function of halo mass; underdense haloes are systematically more triaxial (smaller mean $c/a$) at fixed mass. Right: mean $c/a$ for the full halo population at the six output redshifts $z=0.1$--$1.5$ (colour-coded); haloes are rounder at lower redshift at fixed mass, and the mass dependence turns over for the most massive haloes at low redshift. Error bars are the standard error on the mean per mass bin.}
    \label{fig:ca}
\end{center}
\end{figure}

As a complement to the projected rms ellipticity discussed in Section~\ref{sec:shape_vs_align}, Figure~\ref{fig:ca} shows the three-dimensional minor-to-major axis ratio $c/a$, obtained from the eigenvalues of the halo inertia tensor. In both panels the plotted quantity is the mean of $c/a$ over the haloes in each mass bin, with error bars giving the standard error on the mean. The left panel compares the overdense, total, and underdense subsamples at $z=0.5$ as a function of halo mass. For display simplicity only $z=0.5$ is shown, but the environmental dependence of the axis ratio is the same across the full range $z=0.1$--$1.5$. At fixed mass, haloes in underdense environments have a systematically smaller $c/a$ (are more triaxial) than haloes in overdense environments, across the full mass range probed; this is fully consistent with their larger projected ellipticity (Table~\ref{tab:aia}; Figure~\ref{fig:erms}) and confirms that the environmental shape difference is intrinsic and three-dimensional rather than an artefact of projection.

The right panel shows $c/a$ for the full halo population at the six output redshifts. At fixed mass, haloes are rounder (larger $c/a$) at lower redshift, reflecting the relaxation of older haloes towards more spherical configurations. The mass dependence, however, evolves with redshift: at high redshift $c/a$ increases monotonically with halo mass, whereas towards low redshift the trend becomes non-monotonic, rising with mass to a maximum before turning over and declining for the most massive haloes, which are the youngest, most recently assembled, and dynamically active systems at each epoch.

\section{Halo intrinsic alignment and halo bias}
\label{app:bias}

\begin{figure}
\begin{center}
	\includegraphics[width=0.75\linewidth]{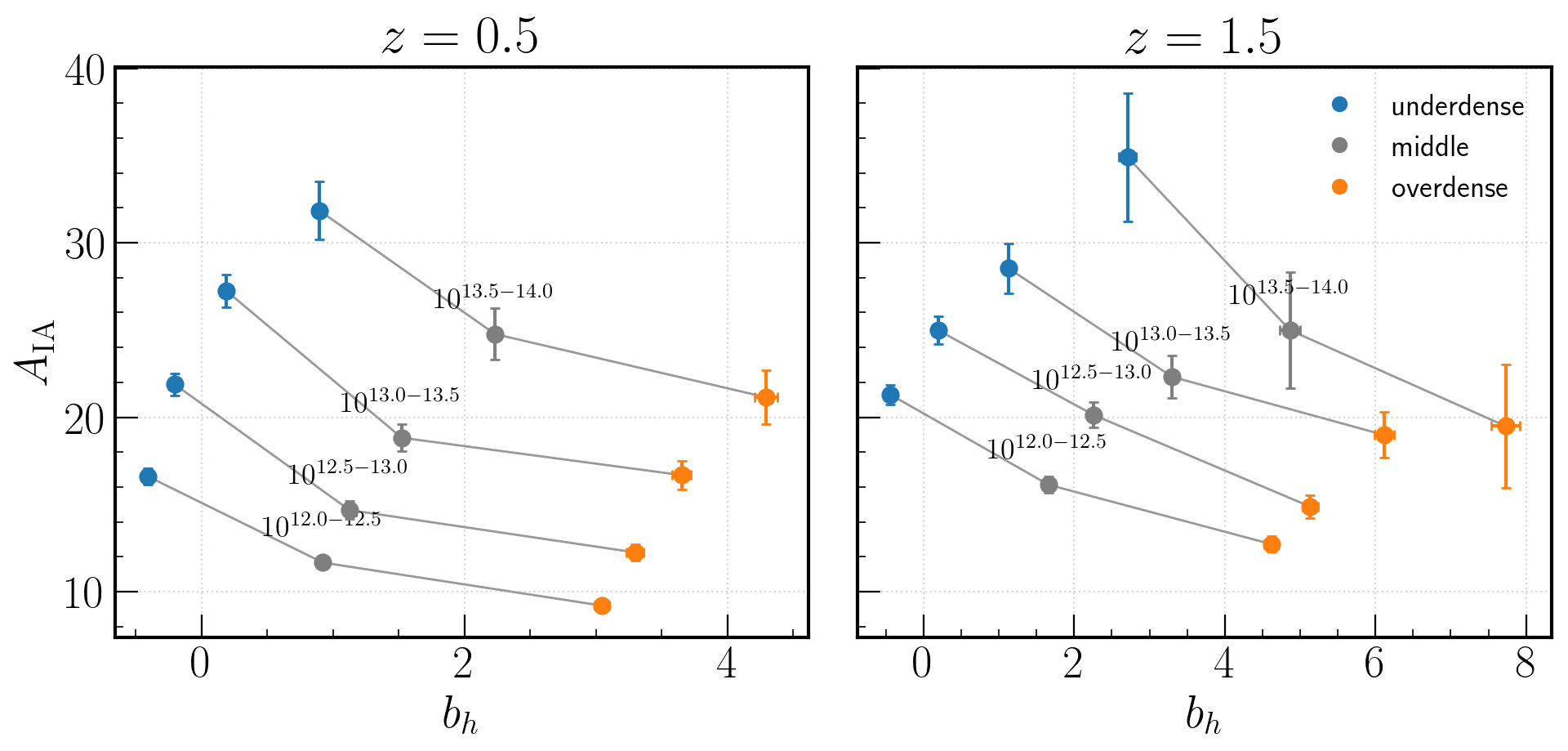}
    \caption{Fitted IA amplitude $A_{\rm IA}$ versus linear halo bias $b_h = P_{\delta h}/P_{\delta\delta}$, at $z=0.5$ (left) and $z=1.5$ (right), for the four narrow mass bins ($10^{12.0}$--$10^{12.5}$ up to $10^{13.5}$--$10^{14}\,\msun$). Colours denote the environment (underdense blue, middledense grey, overdense orange, i.e.\ increasing $\delta_8$); grey lines connect the three environments within each mass bin. At fixed mass, $A_{\rm IA}$ decreases as $b_h$ increases towards denser environments, opposite to the mass trend along which both grow together. Error bars are $1\sigma$ fit uncertainties.}
    \label{fig:aia_bias}
\end{center}
\end{figure}

Figure~\ref{fig:aia_bias} shows the fitted IA amplitude $A_{\rm IA}$ against the linear halo bias $b_h = P_{\delta h}/P_{\delta\delta}$ (Table~\ref{tab:aia}), for the four narrow mass bins and the three environment subsamples (under-, middle-, and overdense) at $z=0.5$ and $z=1.5$; points of the same mass bin are connected. Two trends act in opposite directions. Along the mass sequence (different mass bins at fixed environment), $A_{\rm IA}$ and $b_h$ increase together, recovering the familiar growth of both alignment strength and clustering bias with halo mass. Along the environment sequence (at fixed mass), however, they \emph{anti}-correlate: as $\delta_8$ increases from the under- to the overdense subsample, $b_h$ rises steeply while $A_{\rm IA}$ falls. Strikingly, the underdense subsamples are almost unbiased ($b_h\simeq0$, and even slightly negative for the lowest-mass bin) yet have the \emph{largest} IA amplitude. The IA amplitude is therefore not a single function of the halo bias: at fixed mass it carries a secondary, environmental dependence that runs opposite to the clustering bias -- the ``alignment assembly bias'' discussed in Section~\ref{sec:summary}.

%%%%%%%%%%%%%%%%%%%%%%%%%%%%%%%%%%%%%%%%%%%%%%%%%%

\end{document}